\documentclass[aps,superscriptaddress,floatfix,preprint]{revtex4}
\usepackage{times}
\usepackage{xspace}
\usepackage{graphicx,morefloats}
\usepackage{color}
\usepackage{amsmath, amsthm, amssymb}
\usepackage{multirow}

\newcommand\RMSE{\mathit{RMSE}}
\newcommand\STD{\mathit{STD}}
\begin{document}

\title{Decadal climate predictions using sequential learning algorithms}

\author{Ehud Strobach}
\author{Golan Bel}
\affiliation{Department of Solar Energy and Environmental Physics, Blaustein Institutes for Desert Research, Ben-Gurion University of the Negev, Sede Boqer Campus 84990, Israel}
\begin{abstract}
  Ensembles of climate models are commonly used to improve climate predictions and assess the uncertainties associated with them.
Weighting the models according to their performances holds the promise of further improving their predictions.
Here, we use an ensemble of decadal climate predictions to demonstrate the ability of sequential learning algorithms (SLAs) to reduce the forecast errors and reduce the uncertainties.
Three different SLAs are considered, and their performances are compared with those of an equally weighted ensemble, a linear regression and the climatology. 
Predictions of four different variables--the surface temperature, the zonal and meridional wind, and pressure--are considered. 
The spatial distributions of the performances are presented, and the statistical significance of the improvements achieved by the SLAs is tested.
Based on the performances of the SLAs, we propose one to be highly suitable for the improvement of decadal climate predictions.
\end{abstract}
\maketitle

\section{Introduction}

Global circulation models are the main tools used to simulate future climate conditions. 
There are two main practices by which to initialize these models that represent predictions for two different time scales. 
The first practice corresponds to long-term climate projections.
In this type of simulation, the climate models are initialized in the pre-industrial era (aka uninitialized runs) and integrated forward in time (usually until 2100). 
In these simulations, the atmospheric composition in the past is set according to observations, while for the future, several representative concentration pathways \citep{Moss_2008}, corresponding to different scenarios of atmospheric composition changes, are used. 
These climate simulations are expected to provide information about the response of the climate system to different emission scenarios by predicting the changes in the long-term averages (10 years and more) and the statistics of climate variables, under different atmospheric composition scenarios \citep{Collins2013}. 

The second practice, which is considered in this work, is near-term (decadal) climate predictions intended to provide information on the dynamics of the climate system in time scales shorter than those of significant changes in the atmospheric concentration and the response time of the climate system to such changes. In this practice, the climate models are initialized with observed conditions close to the prediction period. The expected information from these simulations is the dynamics of the monthly to decadal averages of climate variables \citep{Collins15082007,meehl_decadal_2009,Meehl_2013,numerical}, which is of great importance for climate services \citep{Cane_2010}. Recent studies have demonstrated a potential decadal prediction skill in different regions and for different physical processes \citep{Smith_2007,Keenlyside_2008,meehl_decadal_2009,Meehl_2013,pohlmann_2009}.

Despite their relatively short term, decadal climate predictions are still accompanied by large uncertainties, and new methods to improve the predictions and reduce the associated uncertainties are of great interest. One of the main approaches to improving climate predictions is to combine the output from an ensemble of climate models. 
This approach has two known advantages compared with single model predictions. First, it was shown that the ensemble average generates improved predictions \citep{doblas-reyes_2000,doblas-reyes_skill_2003,hagedorn_rationale_2005,palmer_development_2004,Palmer_2000,Kim_2012}; second, the distribution of the ensemble member predictions can provide an estimate of the uncertainties. However, the simple average of climate simulations does not account for the quality differences between the ensemble members; therefore, it is expected that weighting the ensemble members based on their past performances will increase the forecast skill.

Uncertainties in climate predictions can be attributed to three main sources. The first is internal variability, that is, uncertainties due to different initial conditions (either different initialization times or different initialization methods) that were used to run a specific model. The second source is model uncertainties due to different predictions of different models. 
The third source is forcing scenario uncertainties due to different scenarios assumed for the future atmospheric composition \citep{hawkins_potential_2009}. 
The contribution of these sources to the total uncertainty of the climate system varies with the prediction lead time and is also spatially, seasonally and averaging-period dependent \citep{Strobach2015Un}. It was shown that for global and regional decadal climate predictions, scenario uncertainties are negligible compared to the first two sources \citep{hawkins_potential_2009,Cox13072007}.

There are two contributions to the internal variability--variability due to different starting conditions and variability due to different initialization methods. 
Uncertainties due to different starting conditions stem from the chaotic nature of the simulated climate dynamics and cannot be reduced using the ensemble approach. 
However, uncertainties due to different initialization methods and the model variability can be reduced by weighting the members of the ensemble. 
The total reduction of the uncertainty depends on the relative contribution of these sources to the total uncertainty.

Bayesian inference is one of the methods that have been used in the past to weight an ensemble of climate models. 
The main part of this method is the calculation of the posterior density, which is proportional to the product of the prior and the likelihood. 
The Bayesian method optimizes the probability density function (PDF) of the climate variable to the PDF of the data during a learning period and uses it for future predictions. 
It does not assign weights to the climate models; instead, it gives an estimation for the PDF of the predicted climate variable. 
Bayesian inference has been used extensively for projections of future climate \citep{buser_bayesian_2009,buser_bayesian_2010, smith_bayesian_2009,tebaldi_quantifying_2005, tebaldi_use_2007,Furrer_2007,greene_probabilistic_2006, murphy_quantification_2004,Raisanen_2010} and also for near-term climate predictions \citep{Rajagopalan_2002,Robertson_2004}. 
The use of Bayesian inference has reduced the uncertainties of the climate projections and improved their near-term predictions.
However, this method relies on many assumptions regarding the distribution of the climate variables that are not always valid, making the Bayesian inference subjective and variable-dependent.

A second, and more common, method that has been used to improve climate predictions is linear regression \citep{Feng_2011,CHAKRABORTY_2009,doblas-reyes_rationale_2005,fraedrich_combining_1989,kharin_climate_2002, krishnamurti_improved_1999,krishnamurti_multimodel_2000,pavan_multi-model_2000,pena_consolidation_2008,Peng_2002,Yun_2005,yun_improvement_2003}. 
The linear regression method does not assign weights to the ensemble members but rather attempts to find a set of coefficients such that the scalar product of the vector of coefficients and the vector of the model predictions yields the minimal sum of squared errors relative to past observations.
The same set of coefficients is then used to produce future predictions. As a consequence, the regression can be used only for deterministic predictions, that is, the linear combination of the models is calculated to produce better predictions, but there is no straightforward method to estimate the associated uncertainties.
Similarly to the Bayesian method, the regression method also relies on a few inherent assumptions, such as the normal distribution of the prediction errors (therefore, defining the optimal coefficients as those minimizing the sum of squared errors) and the independence of the ensemble member predictions.

Sequential learning algorithms (SLAs, also known as online learning) \citep{cesa2006prediction} weight ensemble members based on their past performances.
These algorithms were shown to improve long-term climate predictions \citep{Monteleoni2010,Monteleoni2011} and seasonal to annual ozone concentration forecasts \citep{Mallet2009,Mallet2010}. More recently, it was shown that decadal climate predictions of the 2m-temperature can be improved using SLAs and can even become skillful when the climatology is added as a member of the ensemble \citep{Strobach2015}. The SLAs have several advantages over the other ensemble methods described above. First, they do not rely on any assumption regarding the models and the distribution of the climate variables. In addition, the weights assigned to the models can be used for model evaluation and the comparison of different parameterization schemes or initialization methods. Third, the weighted ensemble provides not only predictions but also the associated uncertainties. 
All these characteristics suggest that the SLAs are suitable for the improvement of various climate variable predictions.

Here, we test the performances of SLAs in predicting the, previously investigated, 2m-temperature and three additional climate variables--namely, the zonal and meridional components of the surface wind and the surface pressure. The results of the CMIP5 \citep{CMIP5} decadal experiments constitute the ensemble, and the NCEP reanalysis data \citep{kalnay_ncep} are considered as the observations. The performances of the SLAs are compared with those of the regression method. The comparison with the Bayesian method is not straightforward and is not included here.
We also study the effects of different learning periods and different bias correction methods on the SLA performances.
This paper is organized as follows. In Section \ref{sec:Models}, we present the data that we used in this study, including the models and the reanalysis data. In addition, we discuss the different bias correction methods that we used. 
In Section \ref{sec:Methods}, we describe the SLAs and the regression forecasting methods as we implemented them. 
We also provide the details of the climatology that we derived from the reanalysis data. 
In Section \ref{sec:Predictions}, we present the predictions of the different forecasting methods. We also evaluate their global and regional performances based on their root mean square errors ($\RMSE$s). The global and regional uncertainties of the predictions of the different forecasting methods are presented in Section \ref{sec:Uncertainties}. The weights assigned by the SLAs to the different models and to the climatology (all the members of the ensemble) are presented in Section \ref{sec:Weights}. The results are discussed and summarized in Section \ref{sec:Summary}. 

\section{Models and Data\label{sec:Models}}

The decadal experiments were introduced to the Coupled Model Intercomparison Project’s (CMIP) multi-model ensemble in its fifth phase (CMIP5). 
The objective of these experiments is to investigate the ability of climate models to produce skillful future climate predictions for a decadal time scale. 
The climate models in these experiments were initialized with interpolated observation data of the ocean, sea ice and atmospheric conditions, together with the atmospheric composition \citep{CMIP5}. The ability of these simulations to produce skillful predictions was not investigated widely, but it was shown that they can generate skillful predictions in specific regions around the world \citep{Kim_2012,Kirtman_2013,Doblas-Reyes2013,meehl_decadal_2009,pohlmann_2009,Mueller_2012,Meehl_2013,Mueller_2014,Kruschke_2014}.

The CMIP5 decadal experiments were initialized every five years between 1961 and 2011 for 10-year simulations, with three exceptional experiments that were extended to 30-year simulations. 
One of these 30-year experiments was initialized in 1981 and simulated the climate dynamics till 2011. The output of four variables from this experiment is tested here--surface temperature, zonal and meridional surface wind components, and surface pressure. In what follows, we analyze the monthly means of these variables. 

Table \ref{Models_table} shows the eight climate models included in our ensemble. The decadal experiments of the CMIP5 project include a set of runs for each of the models, differing by the starting date and the initialization scheme used. We chose, arbitrarily, the first run of each model. As long as the model variability is the main source of uncertainty, the choice of the realization should not be significant for our analysis. Indeed, it was found that, in the CMIP5 decadal experiments, the model variability is the main source of uncertainty, independent of the prediction lead time, as long as the predictions are not bias corrected. Bias correction reduces mainly the model variability; however, the contribution of the model variability remains important \citep{Strobach2015Un}.

\begin{table}[ht]
\caption{\label{Models_table} Model Availability}
\begin{center}
\begin{tabular}{ p{3cm}  p{3cm}  p{6cm} p{2.5cm}}
 \hline
  Institute ID  & Model Name & Modeling Center (or Group) & Grid (lat X lon)\\
  \hline
  BCC  &   BCC-CSM1.1  &   Beijing Climate Center, China Meteorological Administration & 64 X 128\\
    CCCma & CanCM4 & Canadian Centre for Climate Modelling and Analysis & 64 X 128\\
    CNRM-CERFACS & CNRM-CM5 & Centre National de Recherches Meteorologiques / Centre Europeen de Recherche et Formation Avancees en Calcul Scientifique & 128 X 256\\
    LASG-IAP  & FGOALS-s2  & LASG, Institute of Atmospheric Physics, Chinese Academy of Sciences & 108 X 128\\
    IPSL* & IPSL-CM5A-LR  & Institute Pierre-Simon Laplace & 96 X 96\\
    MIROC & MIROC5 ~~~~~~~~~~~~~~~ MIROC4h &  Atmosphere and Ocean Research Institute (The University of Tokyo), National Institute for Environmental Studies, and Japan Agency for Marine-Earth Science and Technology & 128 X 256 
    
     320 X 640\\
      MRI & MRI-CGCM3  & Meteorological Research Institute & 160 X 320\\
\hline
 \end{tabular}
\begin{flushleft}
{\footnotesize * not available for U and V components of wind}
\end{flushleft}
\end{center}
\end{table}

The NCEP/NCAR reanalysis data \citep{kalnay_ncep} were used as the observation data for the learning and for the evaluation of the forecasting methods’ performances. 
We are aware of other reanalysis projects \citep{ERA_2005,JRA_2007}; however, we selected the NCEP based on its wide use (note that the assessment of the quality of the different reanalysis projects is subjective and is beyond the scope of this paper). The effects of using different reanalysis data are left for future research.

\subsection{Bias correction}\label{sec:bias}

The predictions made by the climate models often suffer from inherent systemic errors \citep{Goddard_2013}, and it is common to apply bias correction methods to the model outputs before analyzing them. For long-term climate projections, this procedure is more straightforward because of the available reference period. Bias correction in decadal climate predictions is not trivial not only because there is no clear reference period but also because some of these experiments are known to have a drift from the initial condition to the model's climatology during the first years of the simulation \citep{meehl_decadal_2009}.

Here, two bias correction methods and the original data were considered. The original data without any bias correction is noted as \textit{no correction}. 
The first bias correction method corresponds to subtracting from each model results their average during the learning period and adding the climatological average (the average of the NCEP/NCAR reanalysis data for the same period). This method is noted as \textit{average correction}. The second bias correction method corresponds to subtracting from each model and for each calendar month the corresponding average during the learning period and adding the NCEP/NCAR reanalysis average for that calendar month during the same learning period. This method is noted as \textit{climatology correction}. 
The two bias correction methods described above do not account for the explicit time dependence of the bias. However, it is reasonable to assume that for decadal climate predictions, the bias does not change considerably with time.

\section{Forecasting methods\label{sec:Methods}} 

In this work, we consider three sequential learning algorithms (SLAs), introduced below. 
More thorough descriptions of the SLAs can be found in Ref. \cite{cesa2006prediction} and in Ref. \cite{Monteleoni2003}. 
We also consider the linear regression (REG) \citep{krishnamurti_multimodel_2000} method in order to compare the performances of the SLAs to the well-known regression method.
The climatology (CLM) is considered here as the threshold for skillful predictions. 
For clarity, the equations that describe the forecasting methods omit the spatial indices. 
However, the forecasting schemes were applied to each of the grid cells independently, thereby allowing the spatial distribution of the weights (or the coefficients in the case of the REG) and the reference climatology. 

\subsection{The EWA and the EGA}

The SLAs use an ensemble of \textit{experts} (climate models), each of which provides a prediction for a future value of a climate variable, to provide a forecast of the climate variable in terms of the weighted average of the ensemble. The process is sequentially repeated with the weights of the models being updated, after each measurement, according to their prediction skill.
We divide the period of the model simulations into two parts. The first part is the learning (or training) period whose data is used to update the model weights in the manner described above, and the second part is used for validating and evaluating the \textit{forecaster} performance. At the end of the learning period, the learning ends and the weights generated by the SLA in the last learning step are used to weight the predictions of the climate models during the validation period.

The deviation of the prediction of model $E$, $f_{E,t}$, from the observed value, $y_t$, determines the \textit{loss function}, $l(f_{E,t},y_t)$, at time $t$.
Similarly, the loss function of the \textit{forecaster} (the SLA) is determined by the deviation of its prediction, $p_t$, from the observed value at time $t$.
The \textit{loss function} is the metric used to evaluate the models’ performances. In our study, we define the \textit{loss function} as the square of the deviation, namely, $ l(f_{E,t},y_t) \equiv (f_{E,t}-y_t)^2 $ for model $E$ and $ l(p_t,y_t) \equiv (p_t-y_t)^2 $ for the \textit{forecaster}.

The output of the Exponentiated Weighted Average (EWA), the first SLA described here, at time $t$ is the set of the weights of the models in the ensemble:
\begin{equation}
w_{E,t}^{EWA}\equiv \frac{1}{Z_t} \cdot w_{E,t-1}^{EWA} \cdot e^{-\eta \cdot l_{E,t}}
\end{equation}
where $\eta$ is a positive number representing the learning rate of the \textit{forecaster} and $Z_t$ is a normalization factor.
The EWA prediction at time $t$ is defined below:
\begin{equation}
p_{t}^{EWA} \equiv \sum_{E=1}^{N_e} w_{E,t-1}^{EWA}\cdot f_{E,t},
\end{equation}
where $N_e$ is the number of models in the ensemble. 

The second SLA considered here is the Exponentiated Gradient Average (EGA). The EGA assigns the weights according to the following rules:
\begin{equation}
w_{E,t}^{EGA} \equiv \frac{1}{Z_t} \cdot w_{E,t-1}^{EGA} \cdot e^{-\eta \cdot l'_{E,t}},
\end{equation}
where $ l'_{E,t}$ is the gradient of the \textit{forecaster loss function} with respect to the weight of model $E$ at time $t-1$.
The mathematical definition of $ l'_{E,t}$ is provided below:
\begin{equation}
 l'(f_{E,t},p_{t}^{EGA},y_t) \equiv \frac{\partial l(p_{t}^{EGA},y_t)}{\partial w_{E,t-1}^{EGA}} =  2 \cdot (p_t^{EGA}-y_t) \cdot f_{E,t},
\end{equation}
where the prediction of the EGA, $p_{t}^{EGA}$, is defined similarly to the prediction of the EWA:
\begin{equation}
p_{t}^{EGA} \equiv \sum_{E=1}^{N_e} w_{E,t-1}^{EGA}\cdot f_{E,t}.
\end{equation}
An important difference between the EWA and the EGA is the fact that under ideal conditions and stationary time series, the EWA converges to the best model in the ensemble, while the EGA converges to the observations \citep{Strobach2015}. 

Note that for the first learning step, one has to assign initial weights to the models. Without any a priori knowledge of the models’ performances, the natural choice is to assign equal weights to all the models. If the hierarchy of the models is known, it is possible to assign their initial weights accordingly.

The learning rate, $\eta$, was optimized by scanning a wide range of values and using the value that resulted in the minimal $\RMSE$ during the learning period. 
However, we added a restriction that the maximal change in the weight of each of the models, between two learning steps, will be smaller than the weight of each model in an equally weighted ensemble--namely, $1/N_e$. This restriction was added to ensure the stability of the weights. The metric that we used for this optimization is defined below:
\begin{equation}
M \equiv \RMSE \cdot \left( 1+\Theta \left( {\max_{E=1,..,N_e,t=1,..,n}} \frac{\Delta w_{E,t}}{(1/N_e)}-1 \right) \right),
\end{equation}
where $\Theta$ represents the Heaviside theta function, and $\RMSE$ is the root mean squared error of the \textit{forecaster} during the $n$ time steps of the learning period.
The $\RMSE$ for a grid cell $(i,j)$ is conventionally defined.
\begin{equation}
\RMSE(i,j)\equiv\sqrt{\left(1/n\right)\sum\limits_{t=1}^{n}(p_t(i,j)-y_t(i,j))^2}.
\end{equation}
The value of $\eta$ that minimizes $M$ was found using a recursive search within a very wide range of values restricted only by the machine precision. 
The optimization was done for each grid cell separately.

\subsection{The Learn-$\alpha$ algorithm}

The basic form of the EWA was modified to explicitly allow switching between \textit{experts}. This switching improves the performance of the SLA when dealing with nonstationary time series. The fixed-shared algorithm introduced in Ref. \cite{Herbster_1998} is defined by the following rules:
\begin{equation}
w_{E,t+1}^{FSA}= \frac{1}{Z_t} \cdot \sum_{E^{*}=1}^{N_e} w_{E,t}^{FSA} \cdot e^{-\eta \cdot l_{E^{*},n}} \cdot K(E,E^{*}),
\end{equation}
where
\begin{equation}
K(E,E^{*};\alpha) \equiv (1-\alpha) \cdot \delta(E,E^{*}) + \frac{\alpha}{N_e-1} \cdot (1-\delta({E,E^{*}})).
\end{equation}
Here, $\alpha \in [0,1]$ is the switching rate parameter, and $\delta (\cdot,\cdot)$ is the Kronecker delta.

The fixed-share algorithm was extended in Ref. \cite{Monteleoni2003} by also learning the optimal switching rate parameter, $\alpha$. 
This modified SLA is known as the Learn-$\alpha$ algorithm (LAA).
In the LAA, the algorithm scans a range of switching rates, $\alpha_j$,  $j\in 1,...,N_{\alpha}$, and assigns weights to each value of $\alpha_j$ based on a loss per alpha function, $l_{t}\left(\alpha_j\right) \equiv -\log\left( \sum_{E=1}^{N_e} w_{E,t} \left(\alpha_j\right) e^{-l_{E,t}}\right)$.
The weights are updated sequentially for both the switching rate and the \textit{experts}. The updating rule for the weight of a specific value, $\alpha_j$, is provided below:
\begin{equation} 
W_{t}\left(\alpha_j\right)=\frac{1}{Z_t}W_{t-1}\left(\alpha_j\right)e^{-l_{t}\left(\alpha_j\right)}.
\end{equation}
The updating rule for the weight of \textit{expert} $E$, given $\alpha_j$, is provided below:
\begin{equation} 
w_{E,t}^{LAA}\left(\alpha_j\right)=\frac{1}{Z_{t}\left(\alpha_j\right)}\sum_{E^*=1}^{N_e} w_{E^{*},t-1}^{LAA}\left(\alpha_j\right)e^{-l_{E^*,t}}K\left(E,E^*;\alpha_j\right).
\end{equation}
The prediction at time $t$, is the weighted average of the \textit{experts} and the different values of $\alpha$.
\begin{equation}
p_{t}^{LAA} = \sum_{E=1}^{N_e} \sum_{j=1}^{N_\alpha} W_{t-1}\left(\alpha_j\right) \cdot w_{E,t-1}^{LAA}\left(\alpha_j\right) \cdot f_{E,t}.
\end{equation}
Here, we adopted a discretization of $\alpha$ to optimize the LAA performance \citep{Monteleoni2003}. 

\subsection{Regression}

The linear regression algorithm considered here is described in Ref. \cite{krishnamurti_multimodel_2000}. 
In this algorithm, the forecast is a linear combination of the climate model predictions as described below:
\begin{equation}
p_t^{REG} = \overline{y} + \sum_{E=1}^N a_E (f_{E,t} - \overline{f}_{E}).
\end{equation}
Here, $\overline{y}\equiv\left(1/n\right)\sum_{t=1}^ny_t$ is the temporal mean of the observed values during the learning period (similarly, $\overline{f}_{E}$ is the temporal mean value of the predicted values by \textit{expert} $E$ during the learning period), and $a_E$ are the regression coefficients minimizing the sum of squared errors during the learning period, $G$, which is defined below: 
\begin{equation}
G \equiv \sum_{t=1}^{n} (p_t-y_t)^2,
\end{equation}
where $n$ is the number of time steps in the learning period.
The algorithm that we used to minimize $G$ involved the elimination of models that were linearly dependent on the other models in the ensemble.

\subsection{Climatology}

The climatology is defined here as the monthly averages of the observed conditions during the learning period. Namely,
\begin{equation}
C_{m}= \sum_{t=1}^{n_1} y_{t,m}
\end{equation}
where $y_{t,m}$ is the observed value in month $m\in[1,12]$ of year $t$ ($t$ is measured in years from the beginning of the simulations), and $n_1$ is the duration of the learning period in years (for simplicity, we assume here that both the learning and the validation periods span an integer number of years). The twelve months of the climatology were replicated to match the duration of the validation period; that is,
\begin{equation}
CLM_{t,m}=C_m,
\end{equation}
for $t\in[n_1+1,n_1+n_2]$ ($n_2$ is the duration of the validation period in years). 
The climatology is often considered as the threshold for a skillful prediction, i.e., a \textit{forecaster} that outperforms the climatology is considered skillful.

\section{Evaluation metrics\label{sec:Metrics}}

Two main evaluation metrics are used here: the average error, quantified by the $\RMSE$ of each of the \textit{forecasters}, and the variability of the ensemble predictions, characterized by their standard deviation, the $\STD$. The global averages of the $\RMSE$ and the $\STD$ are calculated by weighting each grid cell by the fraction of the earth's surface it spans. The precise details are provided here for clarity. During the validation period, the $\RMSE$ of each \textit{forecaster} was calculated for each grid cell (because all the climate variables studied here are two-dimensional, each grid cell has two indices, $(i,j)$) from the time series of the forecast and the observations. Then, the global area-weighted average of the $RMSE$ ($\RMSE_{GAW}$) was calculated as detailed below:
\begin{equation}
\RMSE_{GAW}\equiv\left(1/A_{Earth}\right)\sum\limits_{i,j}A_{i,j}\RMSE(i,j), 
\end{equation}
where $A_{Earth}$ is the total earth's surface area, and $A(i,j)$ is the area spanned by the $(i,j)$ grid cell.
In what follows, we will present both the spatial distribution of the $\RMSE$ and its global average.

Similarly to the $\RMSE$, the variance of the ensemble predictions was calculated for each of the grid cells at each time point and then averaged over time during the validation period. 
The square root of this temporally averaged variance is what we define here as the $\STD$ of each grid cell. The mathematical definition of the $\STD$ is provided below:
\begin{equation} 
\STD(i,j)\equiv\sqrt{(1/n)\sum\limits_{t=1}^{n}\sum\limits_{E=1}^{N} w_E(i,j)(f_{E,t}(i,j)-p_t(i,j))^2 }.
\end{equation}
The global area-weighted average was then calculated:
\begin{equation} 
\STD_{GAW}\equiv\left(1/A_{Earth}\right)\sum\limits_{i,j}A_{i,j}\STD(i,j).
\end{equation}

The skill of the \textit{forecasters} was measured by comparing their $\RMSE$ and $\STD$ to those of some other reference \textit{forecaster}.
For convenience, we define below the $\RMSE$ skill score, $R_{ref,fct}$:
\begin{equation} 
R_{ref,fct} \equiv \frac{\RMSE_{ref}-\RMSE_{fct}}{\frac{1}{2}\left(\RMSE_{ref}+\RMSE_{fct}\right)}.
\end{equation}
The indices $ref$ and $fct$ are used to identify the \textit{forecasters} whose skills are compared. 
Similarly, we define below the $\STD$ skill score, $S_{ref,fct}$:
\begin{equation} 
S_{ref,fct} \equiv \frac{\STD_{ref}-\STD_{fct}}{\frac{1}{2}\left(\STD_{ref}+\STD_{fct}\right)}.
\end{equation}
Unless otherwise specified, we used the climatology as the reference \textit{forecaster} for $R_{ref,fct}$ and the equally weighted ensemble as the reference \textit{forecaster} for $S_{ref,fct}$.
Note that the skill scores are defined such that a \textit{forecaster} with a smaller $\RMSE$ than the reference \textit{forecaster} has a positive $R_{ref,fct}$ score, and similarly, a \textit{forecaster} with a smaller $\STD$ (i.e., smaller uncertainty) than the reference \textit{forecaster} has a positive $S_{ref,fct}$ score.

\section{Predictions\label{sec:Predictions}}

\subsection{Global}

\begin{figure}[t]
\includegraphics[width=\linewidth]{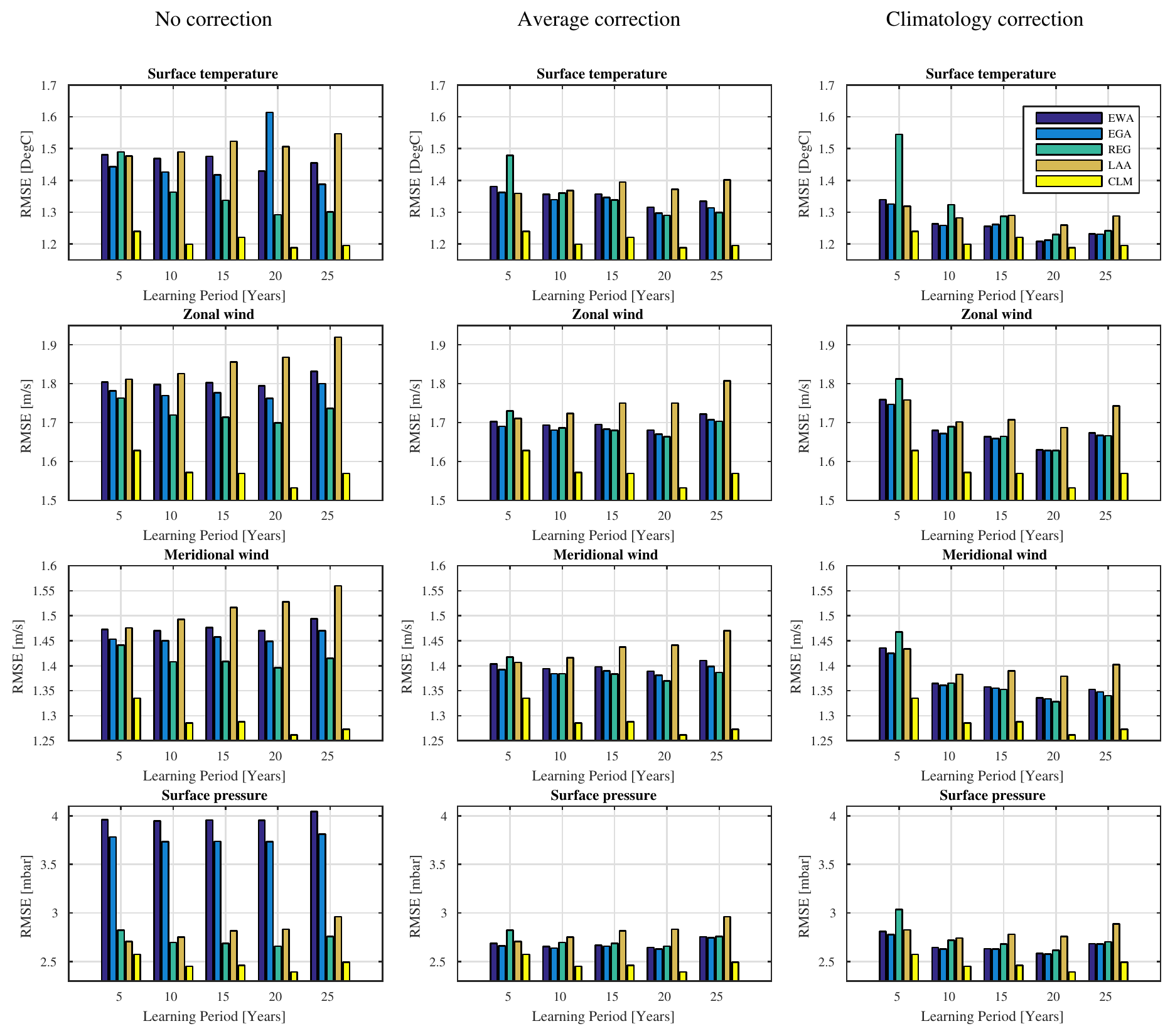}
\caption{\label{fig:RMS_bias_nclm} Globally averaged $\RMSE$. $\RMSE_{GAW}$ for the five forecasting methods (EWA, EGA, REG, LAA and CLM), learning periods of $5$, $10$, $15$, $20$ and $25$ years and the four climate variables (surface temperature, two wind components and pressure). The ensemble used by the \textit{forecasters} does not include the climatology. The left panels correspond to no bias correction, the middle panels correspond to average bias correction, and the right panels correspond to climatology bias correction (see Section \ref{sec:bias} for the details of the different bias correction methods).}
\end{figure}

The simplest measure of the performance of the \textit{forecasters} is the global average of the root mean squared error, $\RMSE_{GAW}$.
Figure \ref{fig:RMS_bias_nclm} shows the $\RMSE_{GAW}$ of the validation period for the five different \textit{forecasters}, EWA, EGA, REG, LAA and CLM, and the different learning periods.
The rows (from top to bottom) correspond to the surface temperature, zonal wind, meridional wind, and pressure, respectively.
The columns (from left to right) correspond to no bias correction, average bias correction, and climatology bias correction, respectively.
The data is provided in Tables 1-4 of the Supplementary Information. The decadal climate simulations considered here span a 30-year period that is split such that the first part is used for learning and the second part is used for the evaluation of the performances; that is, for the five-year learning period, the validation period is the next 25 years, and for the 10-year learning period, the validation period is the next 20 years, etc. The $\RMSE_{GAW}$s of the individual models are not presented because they are much higher than those of the \textit{forecasters}. 
The $\RMSE_{GAW}$ of the equally weighted ensemble is much lower than those of the models, but it is also too high to be included within the scale shown in Fig. \ref{fig:RMS_bias_nclm}.
The bias correction that resulted in the smallest $\RMSE_{GAW}$s is the \textit{climatology correction}, which is described in Section \ref{sec:bias}.

Figure \ref{fig:RMS_bias_nclm} shows that the climatology outperforms all the other \textit{forecasters}, for all the learning periods and bias correction methods studied here. 
Therefore, we added the climatology as an \textit{expert} to the ensemble. Unless otherwise specified, the following results were derived from an ensemble including the climatology as an additional \textit{expert} \citep{Strobach2015}.

\begin{figure}[t]
\includegraphics[width=\linewidth]{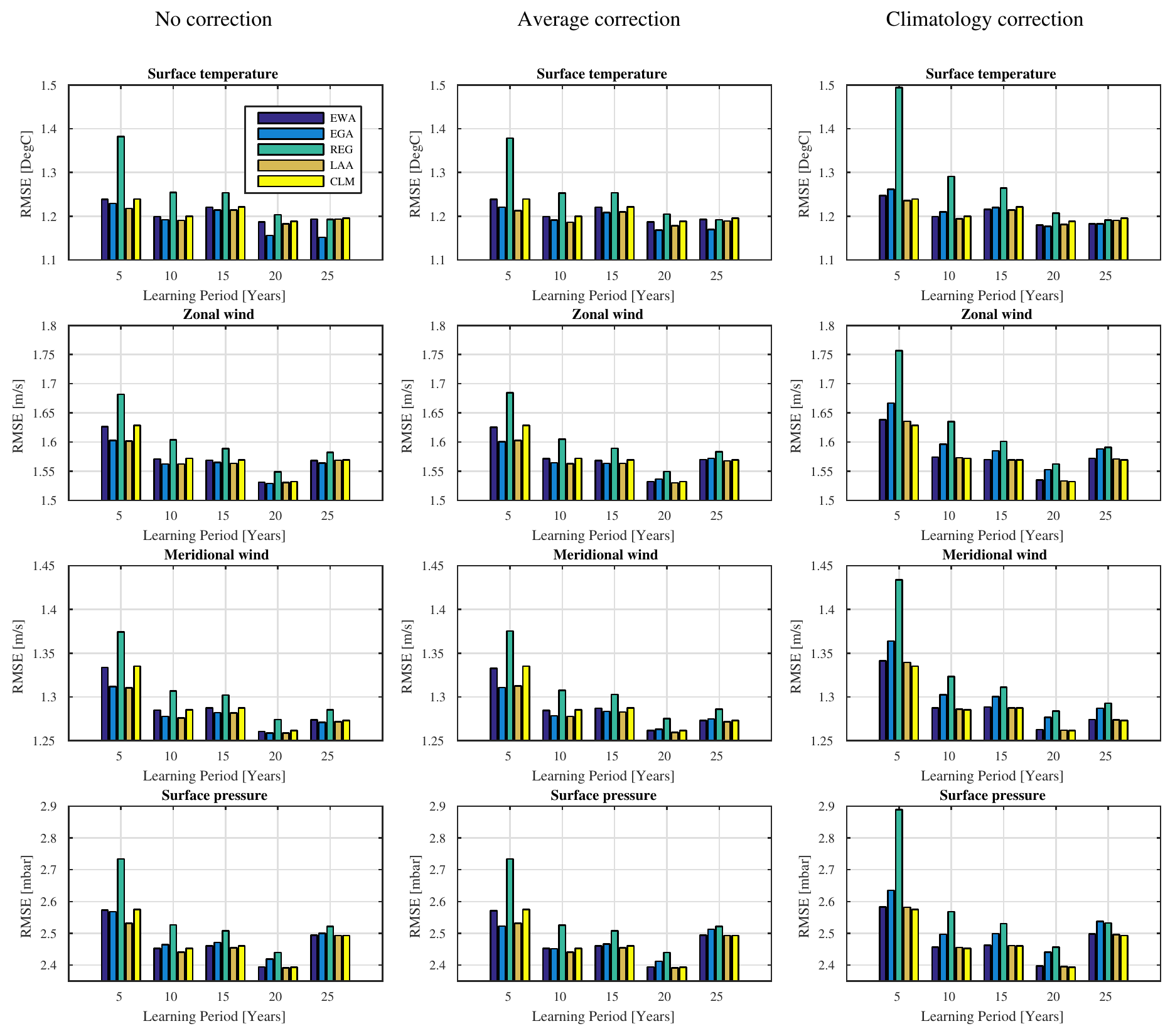}
\caption{\label{fig:RMS_bias} Globally averaged $\RMSE$ with climatology. $\RMSE_{GAW}$ for the five forecasting methods (EWA, EGA, REG, LAA and CLM), learning periods of $5$, $10$, $15$, $20$ and $25$ years and the four climate variables (surface temperature, two wind components and pressure). The ensemble used by the \textit{forecasters} includes the climatology. The left panels correspond to no bias correction, the middle panels correspond to average bias correction, and the right panels correspond to climatology bias correction.}
\end{figure}

Figure \ref{fig:RMS_bias} shows the same results as Figure \ref{fig:RMS_bias_nclm} for an ensemble that includes the climatology. 
In addition, the initial weight assigned to the climatology was $0.5$, whereas the initial weight of all the other models was $0.5/(N_e-1)$ ($N_e-1$ is the number of the models excluding the climatology). This higher initial weight of the climatology was motivated by its superior performance (as shown in Fig. \ref{fig:RMS_bias_nclm} and \citep{Strobach2015}).
The data that was used to generate Fig. \ref{fig:RMS_bias} is provided in Tables 5-8 of the Supplementary Information.

The results of Fig. \ref{fig:RMS_bias} show that the best predictions are obtained using $20$ years of learning and different bias correction methods for different variables and different \textit{forecasters}. The fact that the $\RMSE_{GAW}$ is minimized after $20$ years of learning can be related to two factors: i) for short learning periods, there is a longer prediction period and, therefore, a larger $\RMSE_{GAW}$; ii) for the $25$-year learning period, the time lead from the initialization to the prediction period is long, and in addition, the short five-year prediction period does not represent the climate variability over a time scale of $25$ years (the duration of the learning period).
The $20$ years of learning also ensures that the learning period extends well beyond the drift of the models.
In Table \ref{table:bias}, we detail the bias correction that resulted in the smallest $\RMSE_{GAW}$ for each \textit{forecaster} and for each climate variable. In what follows, we will present only the results of these bias corrections and $20$ years of learning.
We find that all the SLAs have a lower or equal $\RMSE_{GAW}$ than the climatology for the surface temperature and wind components. 
For the surface pressure, only the LAA outperforms the climatology. We also see that, for most climate variables, the $\RMSE_{GAW}$s of the EWA and the climatology are almost equal. 
This is not a coincidence; it reflects the fact that the EWA tracks the best model, which in most grid cells, is the climatology. 
The two other SLAs reduce the $\RMSE_{GAW}$ below that of the climatology by extracting information from the other models in the ensemble. 
The LAA outperforms the EGA for short learning periods ($<15$ years) and for all learning periods in the predictions of the surface pressure. 
This better performance can be attributed to the design of the LAA for the learning of nonstationary data.
The poorer performance, relative to the climatology, of most of the \textit{forecasters} (except for the LAA) in the prediction of the surface pressure is not fully understood. 
However, we found that for the surface pressure, the variability between the models is often larger than its seasonal variability, while all the other climate variables considered here show seasonal variabilities that are larger than the variabilities between the models. 
It is also possible that the model predictions of the monthly mean surface pressure are worse than the predictions of the other climate variables.

\begin{table}[ht]
\caption{\label{table:bias} The optimal bias correction for each \textit{forecaster} and each climate variable. $T$, $U$, $V$, and $P$ denote the surface temperature, zonal wind, meridional wind and pressure, respectively. \textit{nbias}, \textit{bias} and \textit{mbias} correspond to the \textit{no correction}, \textit{average correction} and \textit{climatology correction}, respectively.}
\begin{center}
\begin{tabular}{|c|c|c|c|c|c}
\cline{1-5}
Forecaster & \multicolumn{4}{ c| }{Climate variable} & \\ \cline{1-5}
  & T & U & V & P &\\ \cline{1-5}
 EGA &\textit{nbias} & \textit{nbias}& \textit{nbias}& \textit{bias}& \\ \cline{1-5}
 EWA &\textit{mbias} & \textit{nbias}& \textit{nbias}& \textit{nbias}& \\ \cline{1-5}
 LAA &\textit{bias} & \textit{bias}& \textit{nbias}& \textit{bias}& \\ \cline{1-5}
 REG &\textit{nbias} & \textit{nbias}& \textit{nbias}& \textit{nbias}& \\ \cline{1-5}
 AVG &\textit{mbias} & \textit{mbias}& \textit{mbias}& \textit{mbias}& \\ \cline{1-5}
\end{tabular}
\end{center}
\end{table}

\subsection{Regional}

The $\RMSE_{GAW}$ is convenient because it aims to quantify the performances of the \textit{forecasters} using only one number. 
However, often the more scientifically and practically relevant information are the spatial distributions of the $\RMSE$.
In this subsection, the spatial distribution of the \textit{forecaster} performances will be investigated using the $R_{ref,fct}$ metric defined above.
This metric will allow us to compare the performances of the different \textit{forecasters} and, in particular, to compare their performances to that of the trivial \textit{forecaster}--the climatology. The statistical significance of the improvement achieved by the \textit{forecasters} was tested by introducing the null hypothesis that the temporal distribution of $R_{ref,fct}$ is symmetric around $0$. Grid cells in which the hypothesis was rejected with a $90\%$ confidence level in favor of a better \textit{forecaster} performance are marked with white dots. Similarly, grid cells in which the hypothesis was rejected in favor of a poorer \textit{forecaster} performance are marked with black dots. Grid cells in which the data does not provide enough evidence to reject the null hypothesis are not marked.

\begin{figure}
\centerline{\includegraphics[width=39pc]{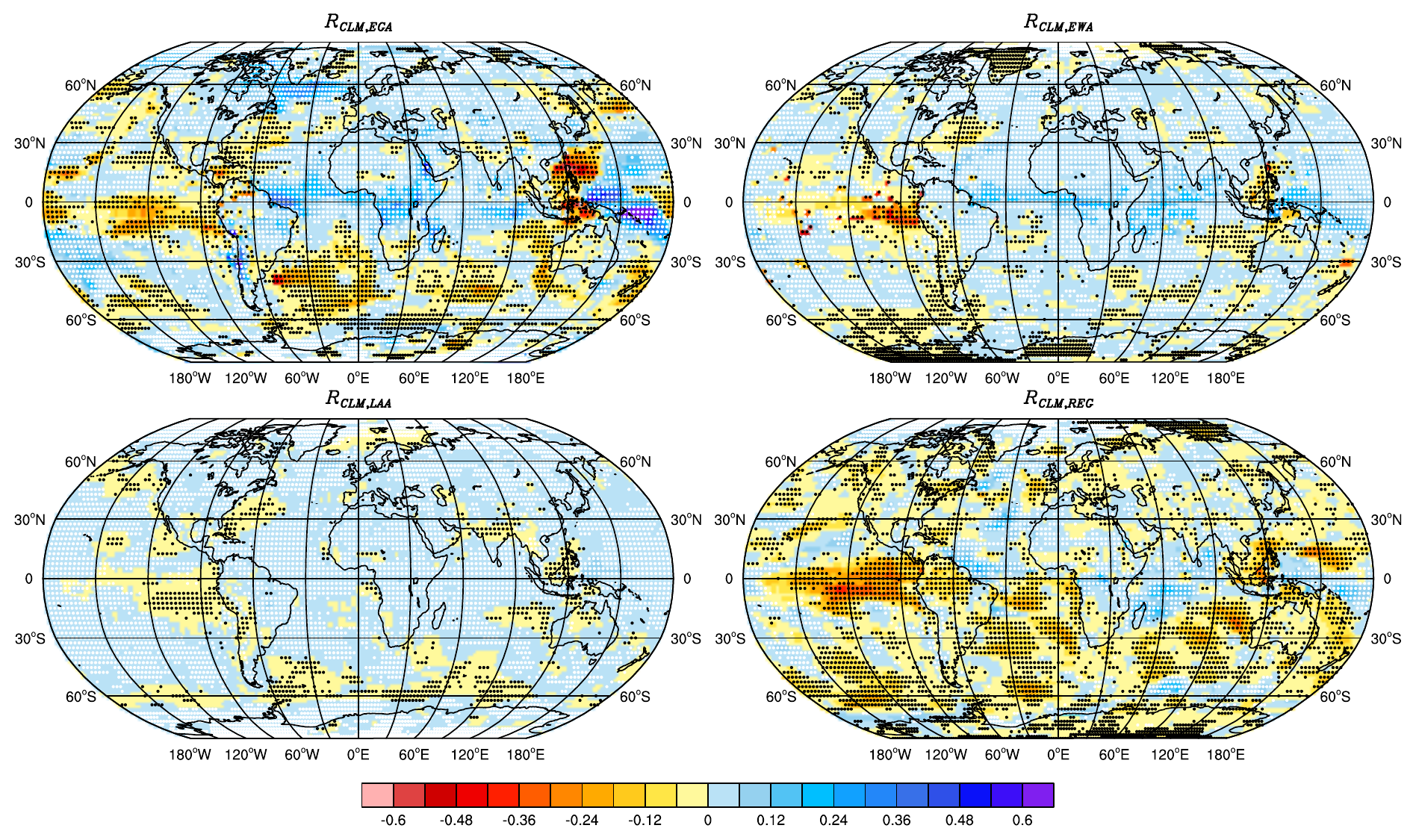}}
\caption{\label{fig:R_tas} Surface temperature $\RMSE$ skill score. Upper left panel: EGA, upper right panel: EWA, lower left panel: LAA and lower right panel: REG. 
Positive values correspond to a smaller $\RMSE$ than the climatology and vice versa. White circles represent significant improvement and black circles represent a significantly poorer performance.}
\end{figure}

Figure \ref{fig:R_tas} depicts the spatial distributions of $R_{CLM,EGA}$ (upper left panel), $R_{CLM,EWA}$ (upper right panel), $R_{CLM,LAA}$ (lower left panel) and $R_{CLM,REG}$ (lower right panel) for the surface temperature. This figure better clarifies the origin of the EGA’s superior performance over the other \textit{forecasters} (as seen from the surface temperature panels, the $20$-year learning period bins of Fig. \ref{fig:RMS_bias}). The largest variability is observed for $R_{CLM,EGA}$ and the smallest variability for $R_{CLM,LAA}$. While the LAA shows a positive skill score over large regions, the score is relatively low, reflecting a small improvement in the prediction compared with the climatology. For the EGA, on the other hand, we see that over regions in the North Atlantic, South America, central Africa, and Oceania, there is a large improvement relative to the climatology, while in regions in the East China Sea, the South Atlantic Ocean and the Eastern Central Pacific Ocean, there is a much poorer performance compared with the climatology. The regression \textit{forecaster} shows a poorer performance compared with the climatology (negative skill score) over most of the globe. 
All the \textit{forecasters} show a positive skill over regions in North Africa, Asia and North America, suggesting that the models are capable of capturing deviations from the climatology in these regions. 

\begin{figure}
\centerline{\includegraphics[width=39pc]{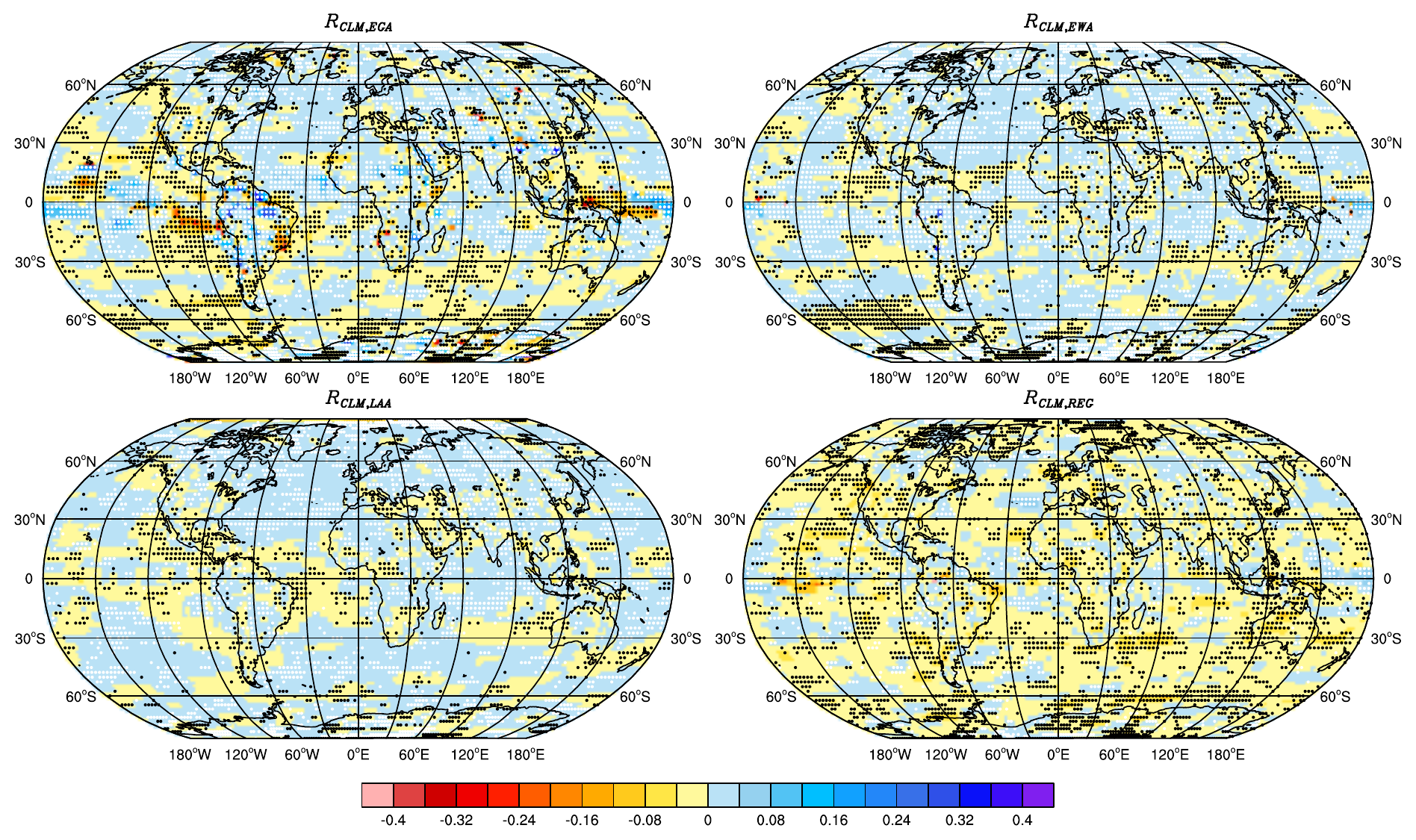}}
\caption{\label{fig:R_uas} Surface zonal wind $\RMSE$ skill score. Upper left panel: EGA, upper right panel: EWA, lower left panel: LAA and lower right panel: REG. 
Positive values correspond to a smaller $\RMSE$ than the climatology and vice versa. White circles represent significant improvement and black circles represent a significantly poorer performance.}
\end{figure}

\begin{figure}
\centerline{\includegraphics[width=39pc]{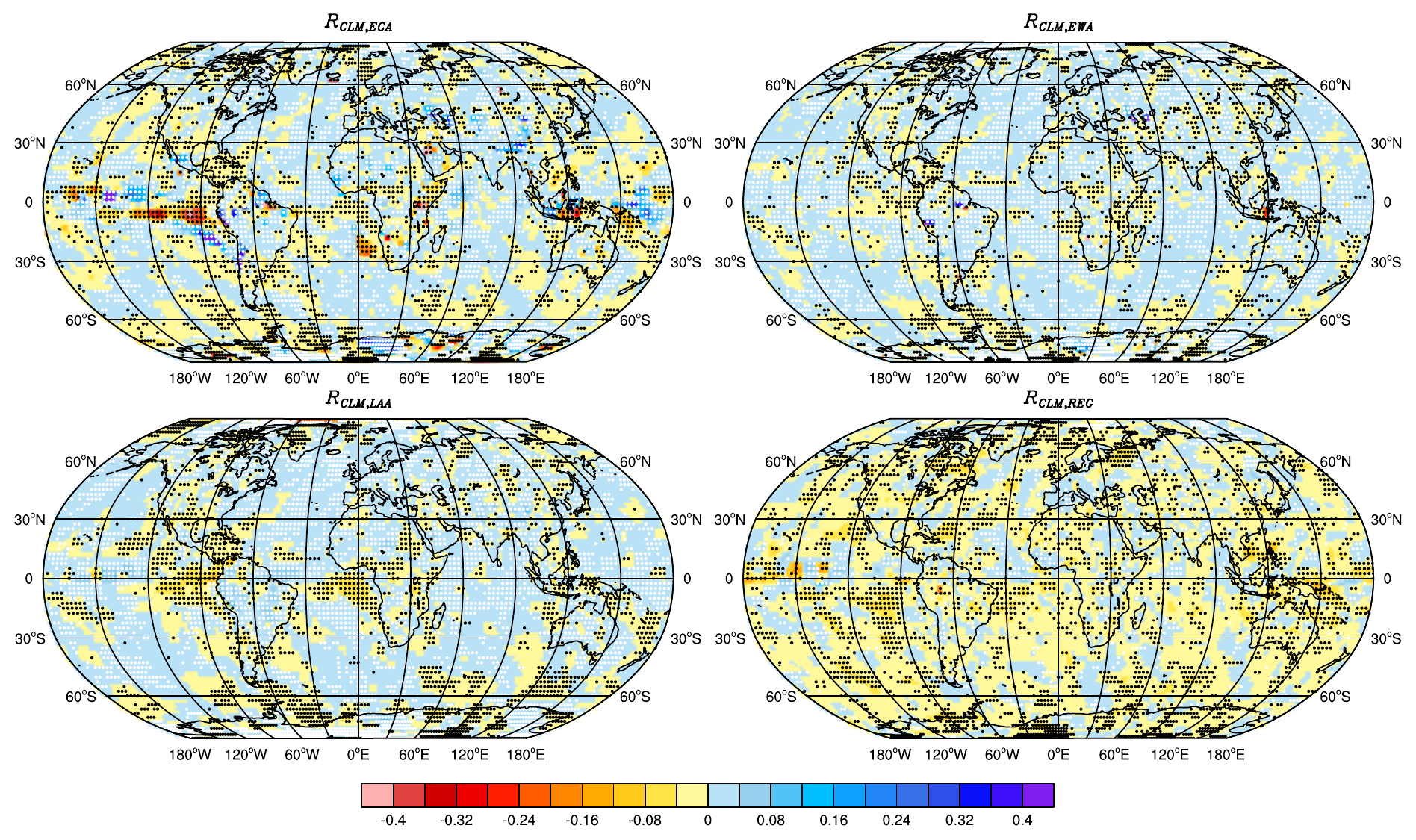}}
\caption{\label{fig:R_vas} Surface meridional wind $\RMSE$ skill score. Upper left panel: EGA, upper right panel: EWA, lower left panel: LAA and lower right panel: REG. 
Positive values correspond to a smaller $\RMSE$ than the climatology and vice versa. White circles represent significant improvement and black circles represent a significantly poorer performance.}
\end{figure}

The spatial distribution of the $\RMSE$ skill score for the zonal and meridional wind components are shown in Figures \ref{fig:R_uas} and \ref{fig:R_vas}, respectively.
Both wind components have similar characteristics. The EGA shows a similar distribution of the skill for the wind components to that found for the surface temperature.
The EWA and the LAA show almost zero skill over most of the globe due to the fact that they both assign a very high weight to the climatology and a very small weight to the other models. Although the improvement relative to the climatology is small, it was found to be statistically significant in many regions. The REG shows a poorer performance compared with the climatology over most of the globe.

\begin{figure}
\centerline{\includegraphics[width=39pc]{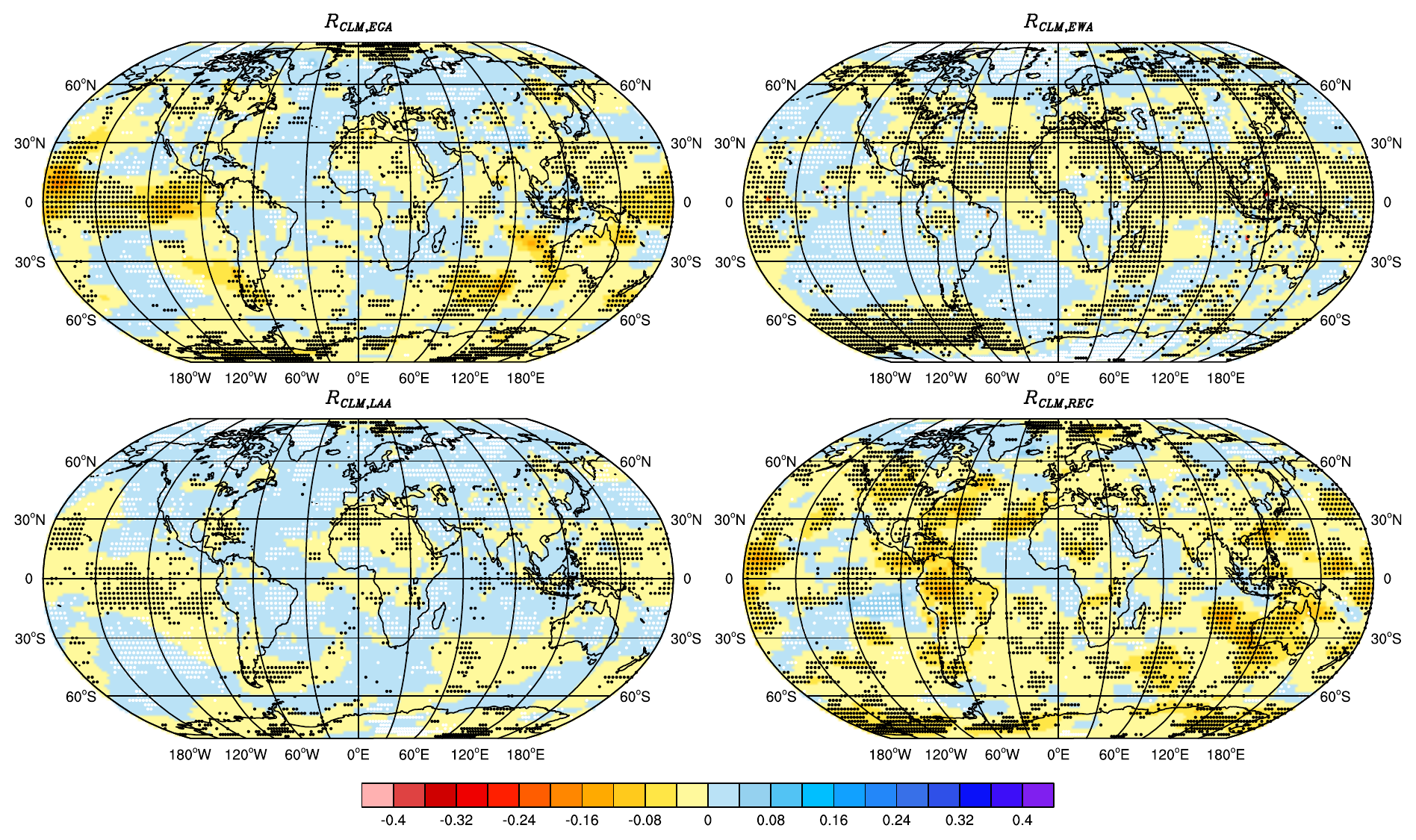}}
\caption{\label{fig:R_ps} Surface pressure $\RMSE$ skill score. Upper left panel: EGA, upper right panel: EWA, lower left panel: LAA and lower right panel: REG. 
Positive values correspond to a smaller $\RMSE$ than the climatology and vice versa. White circles represent significant improvement and black circles represent a significantly poorer performance.}
\end{figure}

Figure \ref{fig:R_ps} shows the spatial distribution of the surface pressure $R_{CLM,EGA}$ (upper left panel), $R_{CLM,EWA}$ (upper right panel), $R_{CLM,LAA}$ (lower left panel) and $R_{CLM,REG}$ (lower right panel). The EGA’s performance for the surface pressure is poor compared with its performance for the other variables. Large regions in the Pacific and Indian Oceans show a larger $\RMSE$ of the EGA than the climatology, while in some regions in the Atlantic Ocean, North Euro-Asia, Greenland and the South Pacific the EGA shows a better performance than the climatology.
The EWA and LAA assign a very high weight to the climatology and, therefore, show an $\RMSE$ skill score close to zero. However, the small improvement achieved by the LAA is statistically significant over most of the globe. The REG shows a poorer performance than the climatology over most regions, with some exceptions in the central Atlantic Ocean and the Arabian Peninsula.

The EGA shows the highest $\RMSE$ skill score over most of the globe for the surface temperature and wind components, while the LAA shows the highest score for the surface pressure.
There are several regions (such as the North Atlantic, North Indian Ocean and North Euro-Asia) where the SLAs seem to provide a smaller $\RMSE$ than the climatology.
This suggests that at least some of the models capture processes that result in a deviation from the climatology and that the SLAs are capable of tracking these models.

\section{Uncertainties\label{sec:Uncertainties}}

The $\RMSE$ is an important measure of the quality of the predictions; however, the uncertainties associated with the predictions of the \textit{forecasters} are crucial for a meaningful assessment of the predictions’ quality. The uncertainties are quantified here using the standard deviation of the ensemble. A natural reference for comparing the variance of the ensemble weighted by the \textit{forecasters} is the variance of the equally weighted ensemble that represents no learning.
It was mentioned earlier that the linear regression does not assign weights to the models in the ensemble but rather attempts to find the linear combination of their predictions that minimizes the sum of squared errors. Therefore, in this section, we will compare the uncertainties of the three SLAs and the equally weighted ensemble, denoted here as AVR.
Our analysis proceeds similarly to the analysis of the $\RMSE$; first we present the globally averaged standard deviation, $\STD_{GAW}$, and then we present the spatial distribution of the $\STD$ skill score.

\subsection{Global}

\begin{figure}[t]
\includegraphics[width=\linewidth]{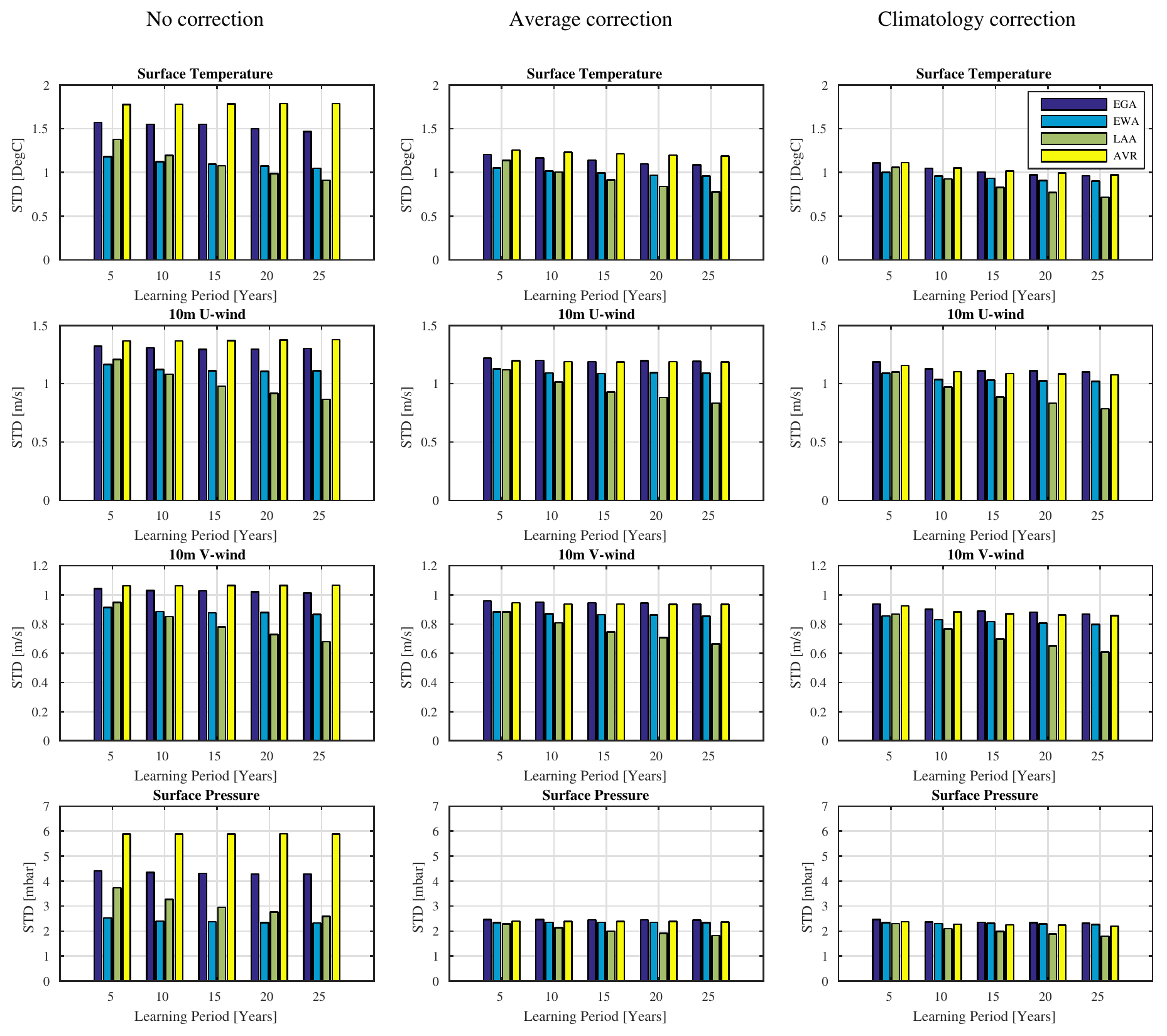}
\caption{\label{fig:STD_bias_nclm} Globally averaged $\STD$. $\STD_{GAW}$ for the three SLAs (EWA, EGA, and LAA) and for the equally weighted ensemble, AVR, for learning periods of $5$, $10$, $15$, $20$ and $25$ years and the four climate variables (surface temperature, two wind components and pressure. The ensemble used by the \textit{forecasters} (and AVR) does not include the climatology. The left panels correspond to no bias correction, the middle panels correspond to average bias correction, and the right panels correspond to climatology bias correction (see Section \ref{sec:bias} for the details of the different bias correction methods).}
\end{figure}

Figure \ref{fig:STD_bias_nclm} shows $\STD_{GAW} $ of the EGA, EWA, LAA and AVR for different learning periods and for the four climate variables considered in this study.
The results of Fig. \ref{fig:STD_bias_nclm} were derived from an ensemble that does not include the climatology. 
The four left panels correspond to no bias correction, the four middle panels correspond to average bias correction and the four right panels correspond to climatology bias correction.
The data is provided in Tables 9-12 of the Supplementary Information.
As expected, the more detailed the bias correction, the smaller the uncertainty because it is associated with the anomaly rather than with the actual prediction. 
We also notice that without bias correction, the EWA has the smallest $\STD_{GAW}$, while with bias correction, the LAA shows the smallest $\STD_{GAW}$. 
Both the EWA and the LAA are expected to have lower $\STD$s because they track the best models. 
With no bias correction, all the SLAs show smaller uncertainties than the equally weighted ensemble, while for the climatology bias correction, the EGA shows a higher $\STD_{GAW}$ than the AVR. This suggests that in large regions, the EGA assigns high weights to models spanning a broad range of predicted values. 
In addition, we notice that the $\STD_{GAW}$ is smaller for longer learning periods, or more precisely, for shorter prediction periods, as expected.
The reduction of $\STD_{GAW}$ is more significant for the LAA because the longer learning allows it to better track the climatology despite the built-in switching rate.

\begin{figure}[t]
\includegraphics[width=\linewidth]{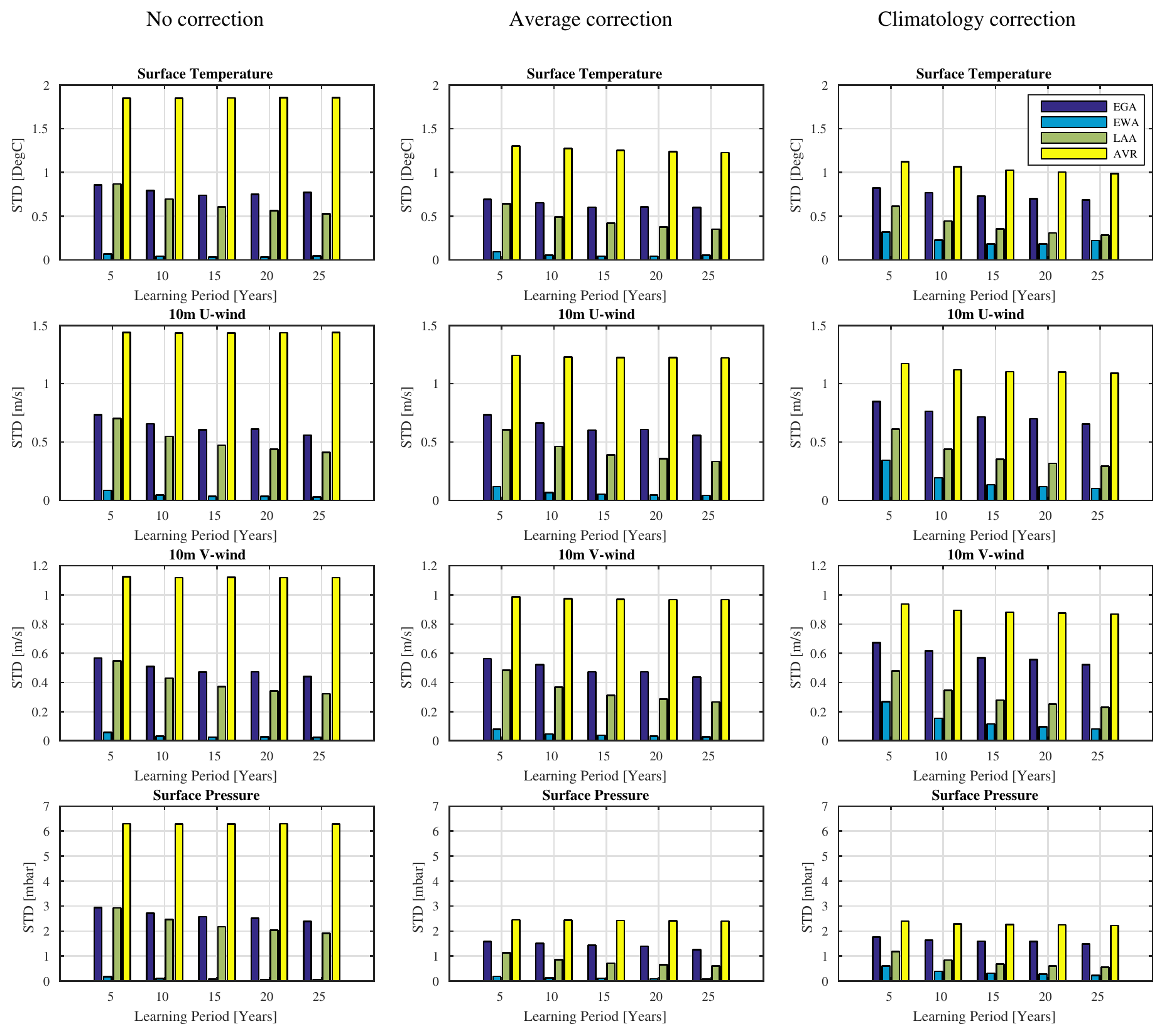}
\caption{\label{fig:STD_bias_clm} Globally averaged $\STD$ with climatology. $\STD_{GAW}$ for the three SLAs (EWA, EGA, and LAA) and for the equally weighted ensemble, AVR, for learning periods of $5$, $10$, $15$, $20$ and $25$ years and the four climate variables (surface temperature, two wind components and pressure). The ensemble used by the \textit{forecasters} (and AVR) includes the climatology. The left panels correspond to no bias correction, the middle panels correspond to average bias correction, and the right panels correspond to climatology bias correction (see Section \ref{sec:bias} for the details of the different bias correction methods).}
\end{figure}

Figure \ref{fig:STD_bias_clm} is similar to Fig. \ref{fig:STD_bias_nclm} but for an ensemble that includes the climatology. 
The data used to generate Fig. \ref{fig:STD_bias_clm} is provided in Tables 13-16 of the Supplementary Information.
It is apparent that in this case, the $\STD_{GAW}$ of all the SLAs is smaller than that of the AVR, and for the longer learning periods, it is much smaller.
The large reduction in the $\STD_{GAW}$ of the EWA and EGA is clearly associated with the fact that they track the climatology in most regions (because it is the best \textit{expert} in these regions). The $\STD_{GAW}$ of the EGA is also reduced because it assigns a high weight to the climatology in many regions, but it still assigns significant weights to the other models; therefore, it has a larger $\STD_{GAW}$ than the other SLAs. The $\STD_{GAW}$ of the equally weighted ensemble does not change much because the climatology is only assigned a weight of $1/N_e$, and in most regions, the climatology is spanned by the other models.

\subsection{Regional}

\begin{figure}
\centering
\centerline{\includegraphics[width=19pc]{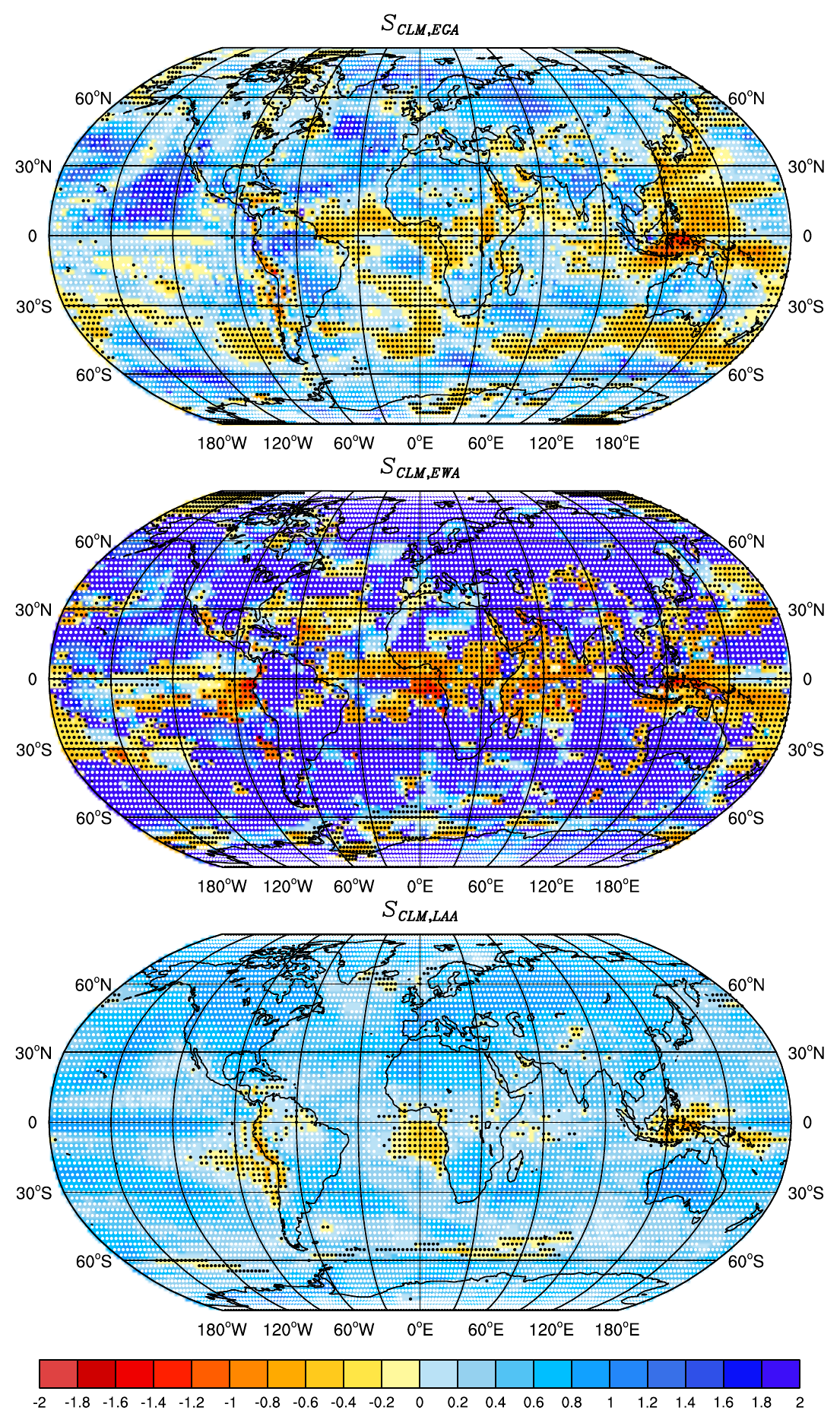}}
\caption{\label{fig:STD_tas} Spatial distribution of the surface temperature $\STD$ skill score. Upper panel: EGA, middle panel: EWA, and lower panel: LAA. 
Positive values correspond to a smaller $\STD$ than the equally weighted ensemble and vice versa. White circles represent a statistically significant reduction of the $\STD$ and black circles represent a statistically significant increase of the $\STD$ relative to the $\STD$ of the equally weighted ensemble.}
\end{figure}

The uncertainty has a large spatial variability. 
We focus on the $20$-year learning period and the ensemble that includes the climatology.
The $\STD$ skill score shows the average temporal variability of the ensemble weighted by the \textit{forecasters} compared with that of the equally weighted ensemble during the validation period. Figure \ref{fig:STD_tas} shows the spatial distribution of the surface temperature $\STD$ skill score for the three SLAs. 
The EGA has a positive $\STD$ skill score (smaller $\STD$ than the equally weighted ensemble) in most of the globe, but there are many regions in which its $\STD$ is significantly larger than that of the AVR. The EWA reduces the $\STD$ over most of the globe except for the tropics. This reduction of the $\STD$ stems from the high weight assigned to the climatology. 
In many regions, the $S_{AVR,EWA}$ is around $2$, which reflects an almost vanishing $\STD$ of the EWA.
The LAA also shows a smaller $\STD$ than the AVR over most of the globe except for small regions in the tropics. 
Similarly to the EWA, the reduction of the uncertainties achieved by the LAA stems from the high weight assigned to the climatology. 
However, one can see that the $S_{AVR,LAA}$ is smaller than the $S_{AVR,EWA}$, which reflects a lower weight of the climatology and higher weights of the other models due to the built-in switching rate in the LAA.

\begin{figure}
\centering
\centerline{\includegraphics[width=19pc]{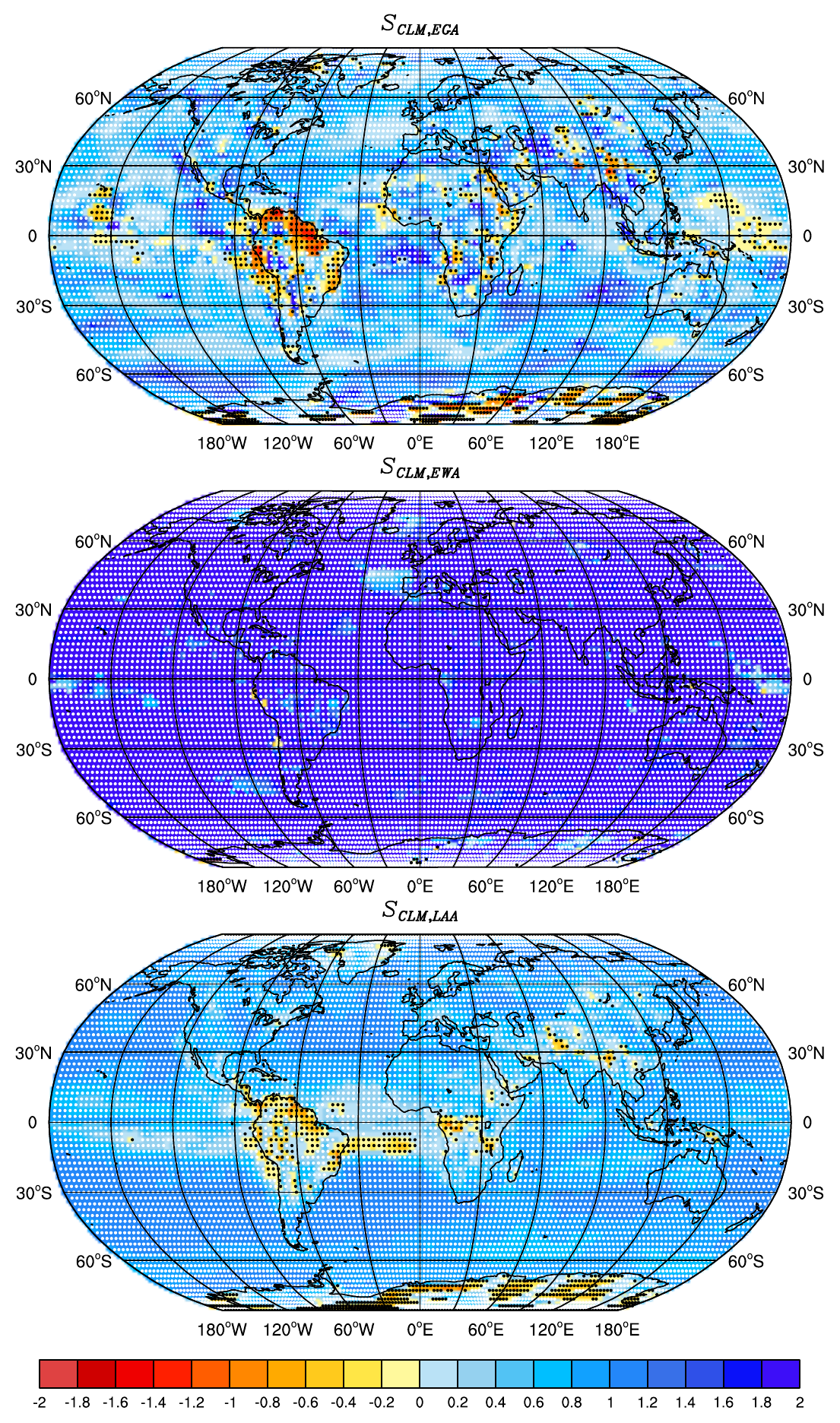}}
\caption{\label{fig:STD_uas} Spatial distribution of the surface zonal wind $\STD$ skill score. Upper panel: EGA, middle panel: EWA, and lower panel: LAA. 
Positive values correspond to a smaller $\STD$ than the equally weighted ensemble and vice versa. White circles represent a statistically significant reduction of the $\STD$ and black circles represent a statistically significant increase of the $\STD$ relative to the $\STD$ of the equally weighted ensemble.}
\end{figure}

\begin{figure}
\centering
\centerline{\includegraphics[width=19pc]{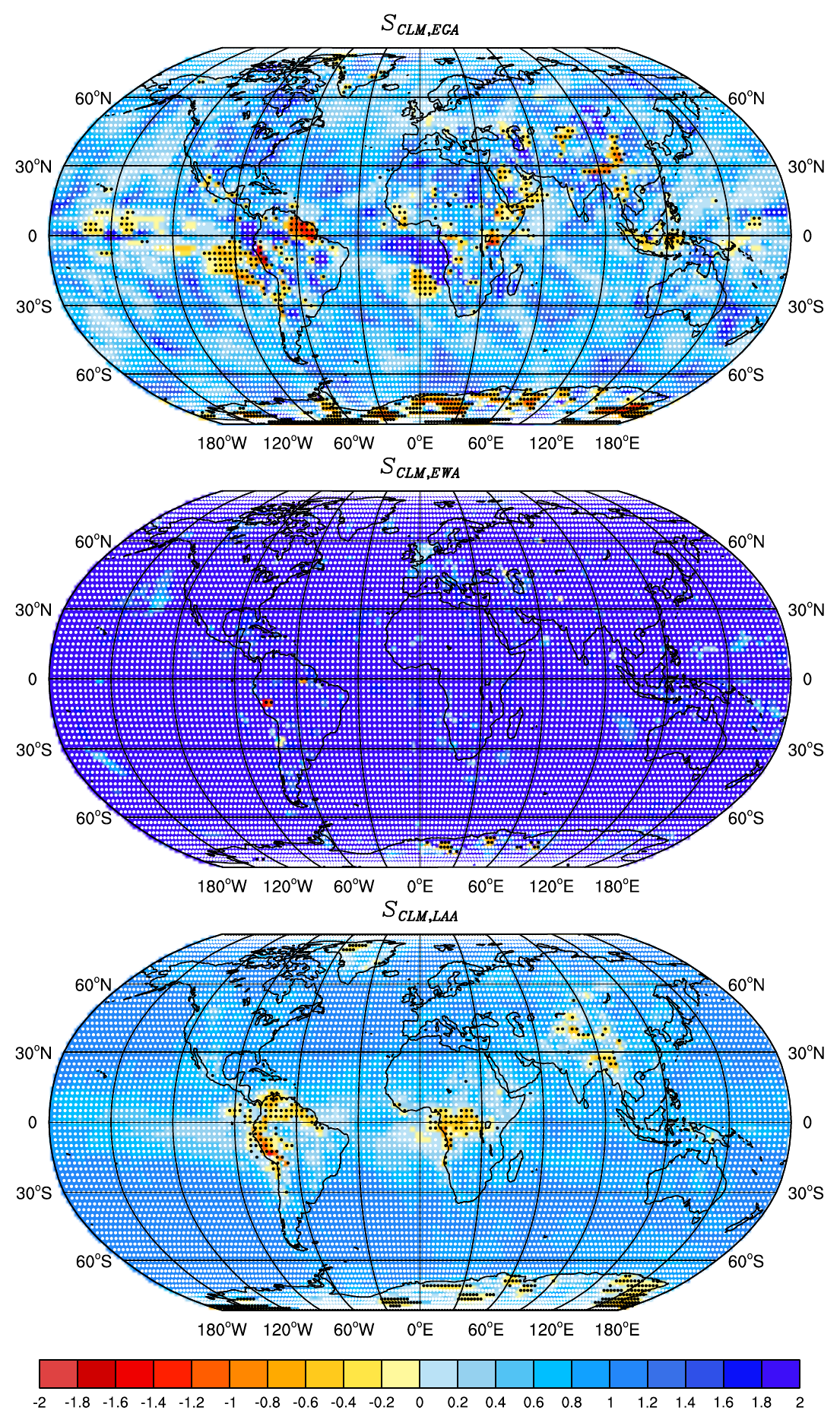}}
\caption{\label{fig:STD_vas} Spatial distribution of the surface meridional wind $\STD$ skill score. Upper panel: EGA, middle panel: EWA, and lower panel: LAA. 
Positive values correspond to a smaller $\STD$ than the equally weighted ensemble and vice versa. White circles represent a statistically significant reduction of the $\STD$ and black circles represent a statistically significant increase of the $\STD$ relative to the $\STD$ of the equally weighted ensemble.}
\end{figure}

Figures \ref{fig:STD_uas} and \ref{fig:STD_vas} show the $\STD$ skill score of the EGA, EWA and LAA for the zonal and meridional wind components.
For the wind components, all the SLAs show significant reductions of the $\STD$ over most of the globe. The EGA and LAA show larger $\STD$s in some small regions in the tropics.
The results suggest that all the SLAs assign a high weight to the climatology, with the EWA almost fully converging to it, while the EGA and LAA extract information from the models as well.

\begin{figure}
\centering
\centerline{\includegraphics[width=19pc]{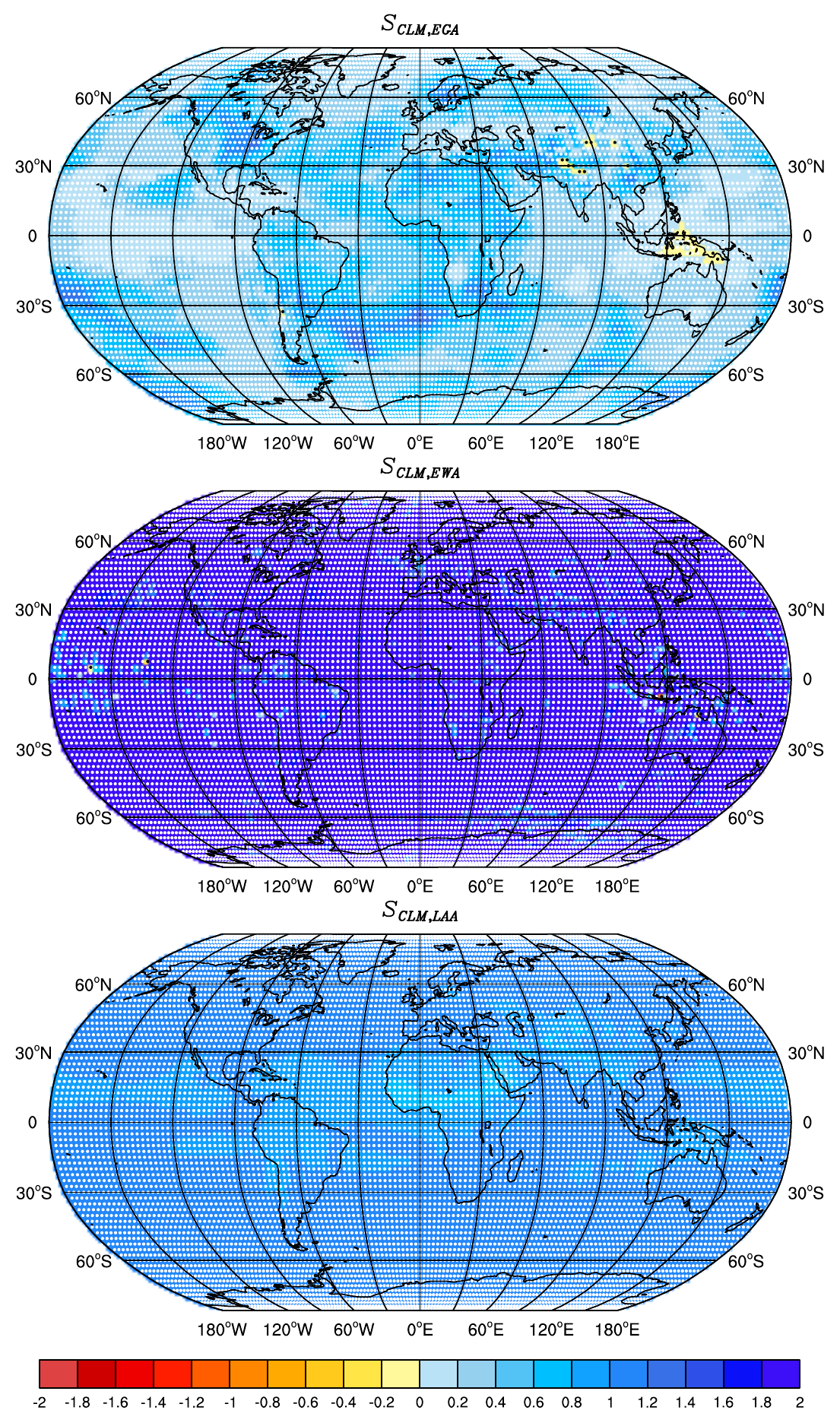}}
\caption{\label{fig:STD_ps} Spatial distribution of the surface pressure $\STD$ skill score. Upper panel: EGA, middle panel: EWA, and lower panel: LAA. 
Positive values correspond to a smaller $\STD$ than the equally weighted ensemble and vice versa. White circles represent a statistically significant reduction of the $\STD$ and black circles represent a statistically significant increase of the $\STD$ relative to the $\STD$ of the equally weighted ensemble.}
\end{figure}

Figure \ref{fig:STD_ps} shows the surface pressure $\STD$ skill score for the three SLAs. 
The EWA and LAA show positive skill scores over the entire globe, and the EGA only shows negative skill in a very small region in Oceania. 
The EWA fully converges to the climatology and has a vanishing $\STD$ (resulting in an $S_{AVR,EWA}$ around $2$ over the entire globe).
The LAA also converges to the climatology, but due to the built-in switching probability, the weight assigned to the climatology is slightly smaller than $1$, and accordingly, the $S_{AVR,LAA}$ is slightly smaller than $2$.

\section{EGA weights\label{sec:Weights}}

\begin{figure}[ht]
\centerline{\includegraphics[width=39pc]{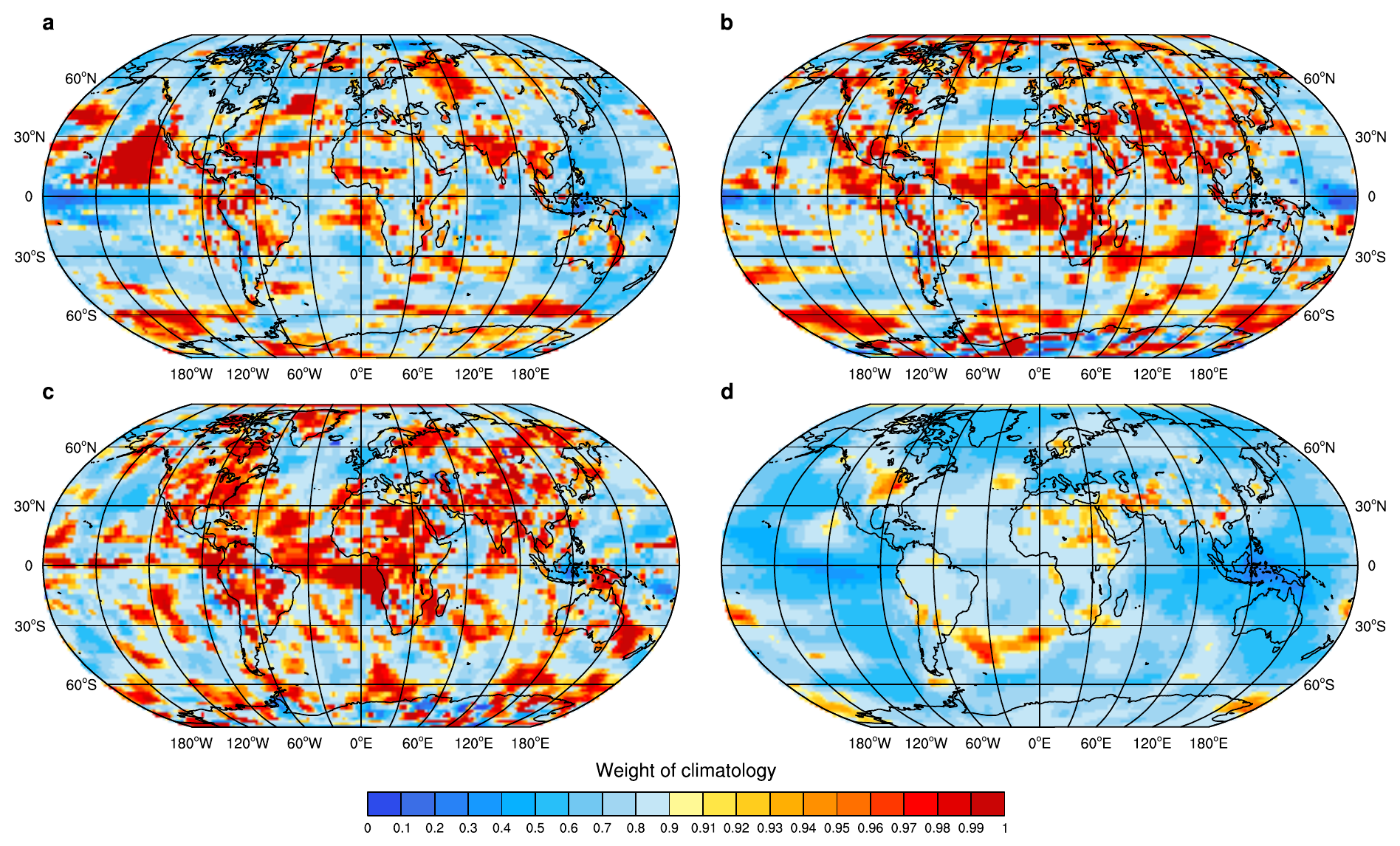}}
\caption{\label{fig:cwega} Spatial distribution of the weight assigned to the climatology by the EGA \textit{forecaster} for \textbf{(a)} surface temperature, \textbf{(b)} surface zonal wind, \textbf{(c)} surface meridional wind and \textbf{(d)} surface pressure.}
\end{figure}

\begin{figure}[ht]
\centerline{\includegraphics[width=39pc]{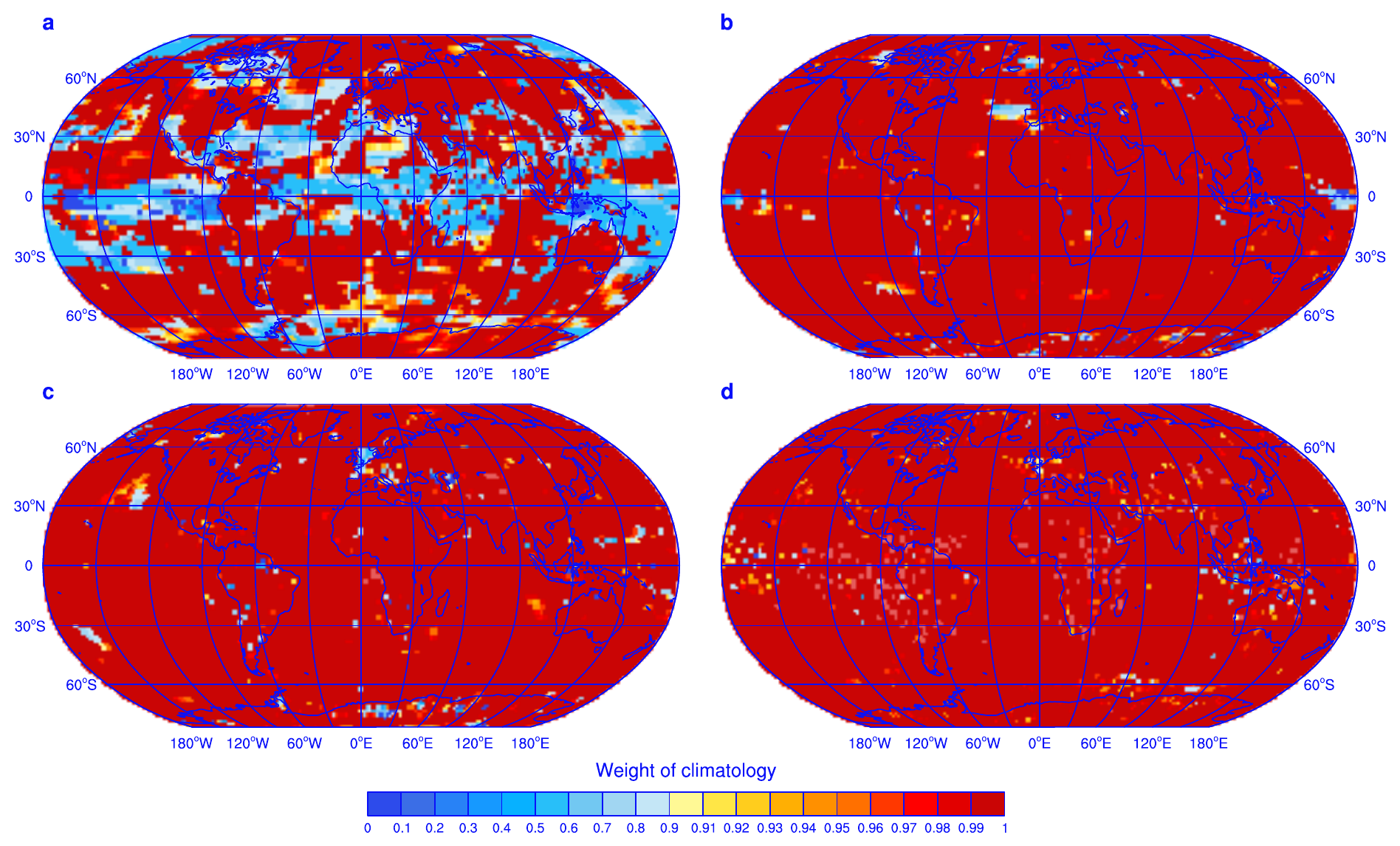}}
\caption{\label{fig:cwewa} Spatial distribution of the weight assigned to the climatology by the EWA \textit{forecaster} for \textbf{(a)} surface temperature, \textbf{(b)} surface zonal wind, \textbf{(c)} surface meridional wind and \textbf{(d)} surface pressure. }
\end{figure}

\begin{figure}[ht]
\centerline{\includegraphics[width=39pc]{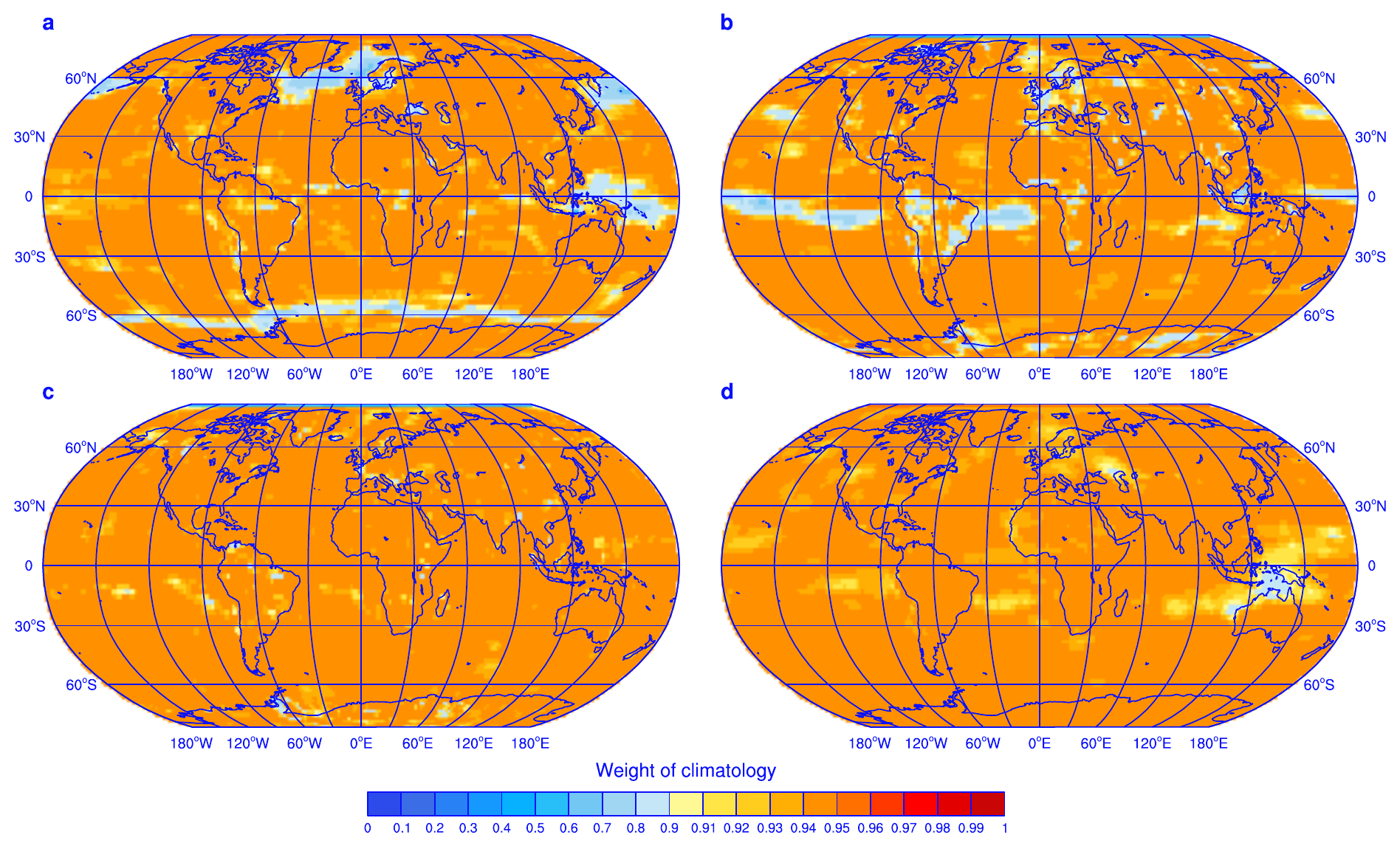}}
\caption{\label{fig:cwlaa} Spatial distribution of the weight assigned to the climatology by the LAA \textit{forecaster} for \textbf{(a)} surface temperature, \textbf{(b)} surface zonal wind, \textbf{(c)} surface meridional wind and \textbf{(d)} surface pressure. }
\end{figure}

Some of the results above regarding the skill of the \textit{forecasters} were explained by the weights assigned to the climatology.
Due to its superior performance, compared with each of the models in the ensemble, it is expected that the SLAs would assign it a high weight.
However, assigning too high a weight to the climatology implies that the \textit{forecaster} is not capable of capturing deviations from the climatology due to the physical processes captured in the models. Ideally, \textit{forecasters} should balance between the smaller $\RMSE$ of the climatology and the additional information available from the other models.

Figures \ref{fig:cwega}, \ref{fig:cwewa} and \ref{fig:cwlaa} show the spatial distribution of the weight assigned to the climatology, for each of the four climate variables, by the EGA, EWA and LAA, respectively. The weights in these figures correspond to the weights assigned at the end of the $20$-year learning period (i.e., the weights used for the predictions).
The colorbar was set to emphasize the differences. The EWA assigns the climatology weights close to $1$ over the entire globe for the surface wind components and pressure.
For the surface temperature, there are large regions in the tropics, close to the North Pole and along the coast of Antarctica where the weight of the climatology is not the only dominant \textit{expert}. Similar patterns are observed for the LAA; however, the weight assigned to the climatology here is never $1$ because this SLA is based on the fixed-share SLA that is designed to have a finite switching probability. Both the weights assigned by the EWA and those assigned by the LAA stem from the fact that these SLAs are designed to track the best \textit{expert}, which in our ensemble turns out to be the climatology over most of the globe.

The EGA assigns a lower weight than the EWA and LAA to the climatology over most of the globe for all the climate variables considered here.
For the surface temperature, only in some regions (mostly in the eastern Pacific Ocean) are the predictions of the EGA dominated by the climatology.
For the surface wind components, the regions dominated by the climatology are somewhat larger.
The weight assigned to the climatology by the EGA for the surface pressure shows a much larger variability (note the nontrivial color map) than the weight assigned for the other variables.
This variability also resulted in a somewhat poorer performance by the EGA in the predictions of this variable. This different performance for the surface pressure may be related to the lower quality of the data for this variable. Unlike the EWA and the LAA, the EGA is not designed to track the best \textit{expert} but rather to track the measurements. Therefore, the lower weight assigned to the climatology suggests that useful information can be extracted from the models, and their ability to capture some of the processes affecting the climate dynamics in decadal time scales can be quantified by the weight assigned to them by the EGA.

\section{Summary and Discussion\label{sec:Summary}}

An ensemble of climate models is known to improve climate predictions and to help better assess the uncertainties associated with them.
In this paper, we tested five different methods to combine the results of the decadal predictions of different models--EWA, EGA, LAA, REG and the equally weighted ensemble. 
The first three \textit{forecasters} represent learning algorithms that weight the ensemble models according to their performances during a learning period.
The REG attempts to find the linear combination of the model predictions that minimizes the sum of squared errors during the learning period, and the equally weighted ensemble represents no learning. We tried different learning periods and found the $20$-year learning experiment to be the most promising. 
This learning period ensures that the learning exceeds well beyond the drift of the models.
The $\RMSE$ and $\STD$ are smaller than those of shorter learning periods, and the results suggest that the lead time (the time from the initialization of the models) has a small effect.
The predictions of the surface temperature, wind and pressure were studied, and their qualities were assessed.

The simple average was shown to have larger errors and larger uncertainties than the \textit{forecasters} that used a learning period to weight/combine the model predictions.
The linear regression showed smaller errors than the equally weighted average. When no bias correction was applied to the data and the ensemble did not include the climatology, the errors of the regression were even smaller than those of the learning algorithms. However, in the more relevant ensemble that includes the climatology, the errors of the linear regression were higher than those of the learning algorithms. This poorer performance is associated with the basic assumptions of the linear regression and its oversimplified method to linearly combine the model predictions. The SLAs do not rely on these assumptions and use more advanced methods to weight the models, resulting in smaller errors.
The REG method does not weight the models but rather finds an optimal linear combination of them; therefore, there is no straightforward method to estimate the uncertainties associated with the linear regression predictions.
The EWA and the LAA were found to be more appropriate in cases in which tracking of the best model is of interest. 
The climatology outperformed all the other models; therefore, the EWA and the LAA converged to it over most of the globe and for all the four climate variables. 
Tracking the best model (by the EWA and LAA) was shown to result in too small uncertainties and thus in overconfident predictions.
For the purpose of improving decadal climate predictions, we found the EGA to be more appropriate because it showed both the ability to reduce the errors and to provide more meaningful estimates of the uncertainties.

Although the globally averaged $\RMSE$ of the EGA is only a few percentage points smaller than that of the climatology, it was shown to be statistically significant.
In addition, we found that in many regions, the improvement is larger. 
The spatial distribution of the EGA performance showed that it is skillful over large continuous regions. This finding suggests that the models were able to capture some physical processes that resulted in deviations from the climatology and that the EGA enabled the extraction of this additional information.
Similarly, the large regions over which the climatology outperforms the \textit{forecasters} may suggest that physical processes, associated with the climate dynamics affecting these regions, are not well captured by the models. The EGA performance was much poorer for the surface pressure than for the other variables. This poorer performance might be related to the quality of the models’ output or to the large fluctuations of this variable. The better predictions of the EWA and LAA for the surface pressure cannot be considered significant because their performance is similar to that of the climatology.
The reduction of the uncertainties is much more substantial than the reduction of the errors and can reach to about $60-70\%$, globally. 
The uncertainties considered here are only those associated with the model variability within the ensemble. The internal uncertainties, scenario uncertainties and other sources of uncertainty were not studied here.

The results presented here are in agreement with previous results (see Ref. \cite{meehl_decadal_2009} and references therein).
However, in this work, monthly means were considered, whereas in previous works, the averages of longer periods, which have smaller fluctuations, were considered. 
A predictive skill of the EGA can be observed in the North Atlantic, in the North Indian Ocean and in some regions in the Pacific Ocean. 
In addition, the EGA showed predictive skill over many land areas, such as North Euro-Asia, Greenland, and, to some extent, also the Americas.
The results suggest that learning algorithms can be used to improve climate predictions and to reduce the uncertainties associated with them.
\acknowledgments

The research leading to these
results has received funding from the European Union Seventh
Framework Programme (FP7/2007-2013) under grant number
[293825].
We acknowledge the World Climate Research Programme's Working Group on Coupled Modelling, which is responsible for the CMIP, and we thank the climate modeling groups (listed in Table \ref{Models_table} of this paper) for producing and making available their model output. For the CMIP, the U.S. Department of Energy's Program for Climate Model Diagnosis and Intercomparison provides coordinating support and led development of software infrastructure in partnership with the Global Organization for Earth System Science Portals.
E.S. wishes to acknowledge a fellowship from the Israel Water Authority.


\bibliographystyle{unsrt}
\bibliography{article}

\appendix
\section{Supplementary Information}

This Supplementary Information provides the globally averaged root mean square errors for the different \textit{forecasters} and different learning periods.
The results provided here were used to select the optimal bias correction method for each \textit{forecaster} and each climate variable. 
In addition, the globally averaged standard deviations of the ensemble weighted by the different \textit{forecasters} are provided.

%
\begin{table}[h]
\caption{\label{table:T_RMSE_nclm} The surface temperature $\RMSE_{GAW}$ for the different \textit{forecasters} and bias correction methods.}
\begin{center}
\begin{tabular}{|c|c|c|c|c|c|c|c}
\cline{1-7}
Forecaster & Bias correction & \multicolumn{5}{ c| }{Learning period} & \\ \cline{1-7}
 &  & 5 & 10 & 15 & 20 & 25 &\\ \cline{1-7}
\multicolumn{1}{ |c  }{\multirow{3}{*}{EGA} } &
\multicolumn{1}{ |c| }{No correction} & 1.4432&	1.4267&	1.4177	&1.3716 &	1.3876 &     \\ \cline{2-7}
\multicolumn{1}{ |c  }{}                        &
\multicolumn{1}{ |c| }{Avg. correction} &1.3624&	1.3394&	1.346&	1.2972&	1.3143 &    \\ \cline{2-7}
\multicolumn{1}{ |c  }{} 			&
\multicolumn{1}{ |c| }{Clm. correction} & 1.3257&	1.2584&	1.2614&	1.2123&	1.2312 &    \\ \cline{1-7}
\multicolumn{1}{ |c  }{\multirow{3}{*}{EWA} } &
\multicolumn{1}{ |c| }{No correction} & 1.4812&	1.4687&	1.4755&	1.4287&	1.4552 &     \\ \cline{2-7}
\multicolumn{1}{ |c  }{}                        &
\multicolumn{1}{ |c| }{Avg. correction} & 1.3811&	1.356&	1.3573&	1.3158&	1.3349 &    \\ \cline{2-7}
\multicolumn{1}{ |c  }{} 			&
\multicolumn{1}{ |c| }{Clm. correction} & 1.3397&	1.2632&	1.2556&	1.2086&	1.2317 &    \\ \cline{1-7}
\multicolumn{1}{ |c  }{\multirow{3}{*}{LAA} } &
\multicolumn{1}{ |c| }{No correction} & 1.4764&	1.4898&	1.5233&	1.5062&	1.5472 &     \\ \cline{2-7}
\multicolumn{1}{ |c  }{}                        &
\multicolumn{1}{ |c| }{Avg. correction} &1.3595&	1.368&	1.3948&	1.3717&	1.4013 &    \\ \cline{2-7}
\multicolumn{1}{ |c  }{} 			&
\multicolumn{1}{ |c| }{Clm. correction} & 1.3193&	1.2825&	1.2906&	1.2593&	1.288 &    \\ \cline{1-7}
\multicolumn{1}{ |c  }{\multirow{3}{*}{REG} } &
\multicolumn{1}{ |c| }{No correction} & 1.4895&	1.3632&	1.3376&	1.2922&	1.301 &     \\ \cline{2-7}
\multicolumn{1}{ |c  }{}                        &
\multicolumn{1}{ |c| }{Avg. correction} & 1.4789&	1.3607&	1.3381&	1.2902&	1.2989 &    \\ \cline{2-7}
\multicolumn{1}{ |c  }{} 			&
\multicolumn{1}{ |c| }{Clm. correction} & 1.5446&	1.3241&	1.2868&	1.2304&	1.2415 &    \\ \cline{1-7}
\multicolumn{1}{ |c  }{\multirow{3}{*}{AVG} } &
\multicolumn{1}{ |c| }{No correction} & 1.7776&	1.7918&	1.8111&	1.8046&	1.8422 &     \\ \cline{2-7}
\multicolumn{1}{ |c  }{}                        &
\multicolumn{1}{ |c| }{Avg. correction} & 1.4152&	1.4118&	1.4184&	1.3904&	1.4194 &    \\ \cline{2-7}
\multicolumn{1}{ |c  }{} 			&
\multicolumn{1}{ |c| }{Clm. correction} & 1.2898&	1.2475&	1.2472&	1.2038&	1.2338 &    \\ \cline{1-7}
CLM & & 1.2393&	1.1994&	1.2213&	1.1881&	1.1956& \\ \cline{1-7}
\end{tabular}
\end{center}
\end{table}

\begin{table}[h]
\caption{\label{table:U_RMSE_nclm} The zonal surface wind $\RMSE_{GAW}$ for the different \textit{forecasters} and bias correction methods.}
\begin{center}
\begin{tabular}{|c|c|c|c|c|c|c|c}
\cline{1-7}
Forecaster & Bias correction & \multicolumn{5}{ c| }{Learning period} & \\ \cline{1-7}
 &  & 5 & 10 & 15 & 20 & 25 &\\ \cline{1-7}
\multicolumn{1}{ |c  }{\multirow{3}{*}{EGA} } &
\multicolumn{1}{ |c| }{No correction}  & 1.7814&	1.7694&	1.7765&	1.7618&	1.8004 &\\ \cline{2-7}
\multicolumn{1}{ |c  }{} &
\multicolumn{1}{ |c| }{Avg. correction} & 1.6905&	1.6806&	1.6829&	1.6698&	1.7076 &\\ \cline{2-7}
\multicolumn{1}{ |c  }{} &
\multicolumn{1}{ |c| }{Clm. correction} & 1.7468&	1.672&	1.6589&	1.628&	1.667 &\\ \cline{1-7}
\multicolumn{1}{ |c  }{\multirow{3}{*}{EWA} } &       
\multicolumn{1}{ |c| }{No correction} & 1.8047&	1.7979&	1.8027&	1.7946&	1.8312 &\\ \cline{2-7}
\multicolumn{1}{ |c  }{} &
\multicolumn{1}{ |c| }{Avg. correction} & 1.7021&	1.6934&	1.695&	1.6806&	1.7221  & \\ \cline{2-7}
\multicolumn{1}{ |c  }{} &
\multicolumn{1}{ |c| }{Clm. correction} & 1.7589&	1.6802&	1.6634&	1.6297&	1.6738 &\\ \cline{1-7}
\multicolumn{1}{ |c  }{\multirow{3}{*}{LAA} } &      
\multicolumn{1}{ |c| }{No correction}  & 1.8114&	1.8259&	1.8557&	1.8678&	1.9199  &\\ \cline{2-7}
\multicolumn{1}{ |c  }{} &
\multicolumn{1}{ |c| }{Avg. correction} & 1.7105&	1.7234&	1.7499&	1.75&	1.8074 &\\ \cline{2-7}
\multicolumn{1}{ |c  }{} &
\multicolumn{1}{ |c| }{Clm. correction} & 1.758&	1.7014&	1.7077&	1.6873&	1.7431 &\\ \cline{1-7}
\multicolumn{1}{ |c  }{\multirow{3}{*}{REG} } &        
\multicolumn{1}{ |c| }{No correction}   & 1.7633&	1.7197&	1.7134&	1.699&	1.7366 &\\ \cline{2-7}
\multicolumn{1}{ |c  }{} &
\multicolumn{1}{ |c| }{Avg. correction} & 1.7296&	1.6865&	1.6801&	1.6637&	1.7034 &\\ \cline{2-7}
\multicolumn{1}{ |c  }{} &
\multicolumn{1}{ |c| }{Clm. correction} & 1.8124&	1.6892&	1.6645&	1.6277&	1.6662    &\\ \cline{1-7}
\multicolumn{1}{ |c  }{\multirow{3}{*}{AVG} } &        
\multicolumn{1}{ |c| }{No correction}  & 1.8947&	1.8982&	1.9094&	1.9126&	1.9665 &\\ \cline{2-7}
\multicolumn{1}{ |c  }{} &
\multicolumn{1}{ |c| }{Avg. correction} & 1.7182&	1.7112&	1.7188&	1.7072&	1.7519 &\\ \cline{2-7}
\multicolumn{1}{ |c  }{} &
\multicolumn{1}{ |c| }{Clm. correction} & 1.7429&	1.6697&	1.6629&	1.6312&	1.677 &\\ \cline{1-7}
CLM        &                 & 1.6285&	1.5719&	1.569&	1.5323&	1.5692 &\\ \cline{1-7}
\end{tabular}
\end{center}
\end{table}

\begin{table}[h]
\caption{\label{table:V_RMSE_nclm} The meridional surface wind $\RMSE_{GAW}$ for the different \textit{forecasters} and bias correction methods.}
\begin{center}
\begin{tabular}{|c|c|c|c|c|c|c|c}
\cline{1-7}
Forecaster & Bias correction & \multicolumn{5}{ c| }{Learning period} & \\ \cline{1-7}
 &  & 5 & 10 & 15 & 20 & 25 &\\ \cline{1-7}
\multicolumn{1}{ |c  }{\multirow{3}{*}{EGA} } &
\multicolumn{1}{ |c| }{No correction}  & 1.4532&	1.4502&	1.4575&	1.4488&	1.4701 &\\ \cline{2-7}
\multicolumn{1}{ |c  }{} &
\multicolumn{1}{ |c| }{Avg. correction} & 1.3923&	1.3839&	1.3897&	1.3808&	1.3987 &\\ \cline{2-7}
\multicolumn{1}{ |c  }{} &
\multicolumn{1}{ |c| }{Clm. correction} & 1.4251&	1.3607&	1.3553&	1.3336&	1.3474 &\\ \cline{1-7}
\multicolumn{1}{ |c  }{\multirow{3}{*}{EWA} } &       
\multicolumn{1}{ |c| }{No correction} & 1.4722&	1.4703&	1.4765&	1.4698&	1.4937 &\\ \cline{2-7}
\multicolumn{1}{ |c  }{} &
\multicolumn{1}{ |c| }{Avg. correction} &1.4033&	1.3941&	1.398&	1.3892&	1.4102 & \\ \cline{2-7}
\multicolumn{1}{ |c  }{} &
\multicolumn{1}{ |c| }{Clm. correction} &1.4355&	1.3647&	1.3577&	1.3355&	1.3528 &\\ \cline{1-7}
\multicolumn{1}{ |c  }{\multirow{3}{*}{LAA} } &      
\multicolumn{1}{ |c| }{No correction}  & 1.4756&	1.493&	1.5169&	1.5275&	1.5599  &\\ \cline{2-7}
\multicolumn{1}{ |c  }{} &
\multicolumn{1}{ |c| }{Avg. correction} &1.4065&	1.4162&	1.4371&	1.4412&	1.47 &\\ \cline{2-7}
\multicolumn{1}{ |c  }{} &
\multicolumn{1}{ |c| }{Clm. correction} & 1.4336&	1.3829&	1.3896&	1.3792&	1.4019 &\\ \cline{1-7}
\multicolumn{1}{ |c  }{\multirow{3}{*}{REG} } &        
\multicolumn{1}{ |c| }{No correction}   & 1.4412&	1.4078&	1.4081&	1.3956&	1.4149  &\\ \cline{2-7}
\multicolumn{1}{ |c  }{} &
\multicolumn{1}{ |c| }{Avg. correction} & 1.4171&	1.3839&	1.3831&	1.3692&	1.3867 &\\ \cline{2-7}
\multicolumn{1}{ |c  }{} &
\multicolumn{1}{ |c| }{Clm. correction} & 1.4673&	1.3651&	1.3526&	1.3279&	1.3401    &\\ \cline{1-7}
\multicolumn{1}{ |c  }{\multirow{3}{*}{AVG} } &        
\multicolumn{1}{ |c| }{No correction}  &1.5319&	1.5349&	1.5424&	1.544&	1.5708 &\\ \cline{2-7}
\multicolumn{1}{ |c  }{} &
\multicolumn{1}{ |c| }{Avg. correction} & 1.4145&	1.4109&	1.4174&	1.4114&	1.4323 &\\ \cline{2-7}
\multicolumn{1}{ |c  }{} &
\multicolumn{1}{ |c| }{Clm. correction} & 1.4185&	1.36&	1.3563&	1.3355&	1.352&\\ \cline{1-7}
CLM        &                 & 1.3352&	1.2853&	1.2876&	1.2617&	1.2731 &\\ \cline{1-7}
\end{tabular}
\end{center}
\end{table}

\begin{table}[h]
\caption{\label{table:P_RMSE_nclm} The surface pressure $\RMSE_{GAW}$ for the different \textit{forecasters} and bias correction methods.}
\begin{center}
\begin{tabular}{|c|c|c|c|c|c|c|c}
\cline{1-7}
Forecaster & Bias correction & \multicolumn{5}{ c| }{Learning period} & \\ \cline{1-7}
 &  & 5 & 10 & 15 & 20 & 25 &\\ \cline{1-7}
\multicolumn{1}{ |c  }{\multirow{3}{*}{EGA} } &
\multicolumn{1}{ |c| }{No correction}  &3.7837&	3.735&	3.7369&	3.7327&	3.8133 &\\ \cline{2-7}
\multicolumn{1}{ |c  }{} &
\multicolumn{1}{ |c| }{Avg. correction} &2.6627&	2.6388&	2.6592&	2.6314&	2.7475 &\\ \cline{2-7}
\multicolumn{1}{ |c  }{} &
\multicolumn{1}{ |c| }{Clm. correction} & 2.7774&	2.628&	2.6278&	2.5765&	2.682&\\ \cline{1-7}
\multicolumn{1}{ |c  }{\multirow{3}{*}{EWA} } &       
\multicolumn{1}{ |c| }{No correction} & 3.9589&	3.9483&	3.9562&	3.9528&	4.043 &\\ \cline{2-7}
\multicolumn{1}{ |c  }{} &
\multicolumn{1}{ |c| }{Avg. correction} &2.689&	2.6561&	2.6687&	2.6458&	2.7556 & \\ \cline{2-7}
\multicolumn{1}{ |c  }{} &
\multicolumn{1}{ |c| }{Clm. correction} & 2.8111&	2.6444&	2.6335&	2.5824&	2.6838 &\\ \cline{1-7}
\multicolumn{1}{ |c  }{\multirow{3}{*}{LAA} } &      
\multicolumn{1}{ |c| }{No correction}  & 4.1878&	4.1653&	4.2056&	4.2313&	4.3497   &\\ \cline{2-7}
\multicolumn{1}{ |c  }{} &
\multicolumn{1}{ |c| }{Avg. correction} & 2.7081&	2.7503&	2.8186&	2.8327&	2.9609 &\\ \cline{2-7}
\multicolumn{1}{ |c  }{} &
\multicolumn{1}{ |c| }{Clm. correction} & 2.8254&	2.7417&	2.7802&	2.7597&	2.8885&\\ \cline{1-7}
\multicolumn{1}{ |c  }{\multirow{3}{*}{REG} } &        
\multicolumn{1}{ |c| }{No correction}   &  2.8237&	2.6976&	2.6875&	2.6583&	2.7572 &\\ \cline{2-7}
\multicolumn{1}{ |c  }{} &
\multicolumn{1}{ |c| }{Avg. correction} & 2.824&	2.6978&	2.6877&	2.6584&	2.7574 &\\ \cline{2-7}
\multicolumn{1}{ |c  }{} &
\multicolumn{1}{ |c| }{Clm. correction} & 3.035&	2.7184&	2.6809&	2.6158&	2.7049    &\\ \cline{1-7}
\multicolumn{1}{ |c  }{\multirow{3}{*}{AVG} } &        
\multicolumn{1}{ |c| }{No correction}  & 6.4757&	6.4749&	6.4933&	6.4998&	6.5855&\\ \cline{2-7}
\multicolumn{1}{ |c  }{} &
\multicolumn{1}{ |c| }{Avg. correction} & 2.6786&	2.659	&2.6839&	2.6639&	2.7813 &\\ \cline{2-7}
\multicolumn{1}{ |c  }{} &
\multicolumn{1}{ |c| }{Clm. correction} & 2.7506&	2.6155&	2.6208&	2.5701&	2.6785&\\ \cline{1-7}
CLM        &                 &2.5746&	2.4531&	2.4606&	2.3935&	2.4933 &\\ \cline{1-7}
\end{tabular}
\end{center}
\end{table}

\begin{table}[h]
\caption{\label{table:T_RMSE} The surface temperature $\RMSE_{GAW}$ for the different \textit{forecasters} and bias correction methods. The climatology is included in the ensemble.}
\begin{center}
\begin{tabular}{|c|c|c|c|c|c|c|c}
\cline{1-7}
Forecaster & Bias correction & \multicolumn{5}{ c| }{Learning period} & \\ \cline{1-7}
 &  & 5 & 10 & 15 & 20 & 25 &\\ \cline{1-7}
\multicolumn{1}{ |c  }{\multirow{3}{*}{EGA} } &
\multicolumn{1}{ |c| }{No correction} & 1.229&	1.1919&	1.2146&	1.1562&	1.1516 &     \\ \cline{2-7}
\multicolumn{1}{ |c  }{}                        &
\multicolumn{1}{ |c| }{Avg. correction} &1.2204&	1.1909&	1.2088&	1.1681&	1.17 &    \\ \cline{2-7}
\multicolumn{1}{ |c  }{} 			&
\multicolumn{1}{ |c| }{Clm. correction} & 1.2616&	1.2095&	1.2202&	1.177&	1.183 &    \\ \cline{1-7}
\multicolumn{1}{ |c  }{\multirow{3}{*}{EWA} } &
\multicolumn{1}{ |c| }{No correction} & 1.2382&	1.1992&	1.2207&	1.1869&	1.1937 &     \\ \cline{2-7}
\multicolumn{1}{ |c  }{}                        &
\multicolumn{1}{ |c| }{Avg. correction} & 1.2389&	1.1993&	1.2205&	1.1873&	1.1926 &    \\ \cline{2-7}
\multicolumn{1}{ |c  }{} 			&
\multicolumn{1}{ |c| }{Clm. correction} & 1.2468&	1.1993&	1.2157&	1.1797&	1.1825 &    \\ \cline{1-7}
\multicolumn{1}{ |c  }{\multirow{3}{*}{LAA} } &
\multicolumn{1}{ |c| }{No correction} & 1.2179&	1.1907&	1.2141&	1.1825&	1.1931 &     \\ \cline{2-7}
\multicolumn{1}{ |c  }{}                        &
\multicolumn{1}{ |c| }{Avg. correction} &1.2125&	1.1865&	1.21&	1.1787&	1.1888 &    \\ \cline{2-7}
\multicolumn{1}{ |c  }{} 			&
\multicolumn{1}{ |c| }{Clm. correction} & 1.2354&	1.1945&	1.2146&	1.1814&	1.1904 &    \\ \cline{1-7}
\multicolumn{1}{ |c  }{\multirow{3}{*}{REG} } &
\multicolumn{1}{ |c| }{No correction} & 1.3822&	1.2546&	1.2539&	1.2035&	1.1928 &     \\ \cline{2-7}
\multicolumn{1}{ |c  }{}                        &
\multicolumn{1}{ |c| }{Avg. correction} & 1.3784&	1.2526&	1.2537&	1.2046&	1.1918 &    \\ \cline{2-7}
\multicolumn{1}{ |c  }{} 			&
\multicolumn{1}{ |c| }{Clm. correction} & 1.4939&	1.291&	1.2645&	1.2073&	1.1915 &    \\ \cline{1-7}
\multicolumn{1}{ |c  }{\multirow{3}{*}{AVG} } &
\multicolumn{1}{ |c| }{No correction} & 1.6645&	1.6783&	1.6983&	1.6896&	1.726 &     \\ \cline{2-7}
\multicolumn{1}{ |c  }{}                        &
\multicolumn{1}{ |c| }{Avg. correction} & 1.3597&	1.3562&	1.3644&	1.3366&	1.3649 &    \\ \cline{2-7}
\multicolumn{1}{ |c  }{} 			&
\multicolumn{1}{ |c| }{Clm. correction} & 1.2748&	1.2328&	1.2342&	1.1913&	1.219 &    \\ \cline{1-7}
CLM & & 1.2393&	1.1994&	1.2213&	1.1881&	1.1956& \\ \cline{1-7}
\end{tabular}
\end{center}
\end{table}

\begin{table}[h]
\caption{\label{table:U_RMSE} The zonal surface wind $\RMSE_{GAW}$ for the different \textit{forecasters} and bias correction methods. The climatology is included in the ensemble.}
\begin{center}
\begin{tabular}{|c|c|c|c|c|c|c|c}
\cline{1-7}
Forecaster & Bias correction & \multicolumn{5}{ c| }{Learning period} & \\ \cline{1-7}
 &  & 5 & 10 & 15 & 20 & 25 &\\ \cline{1-7}
\multicolumn{1}{ |c  }{\multirow{3}{*}{EGA} } &
\multicolumn{1}{ |c| }{No correction}  & 1.6029          & 1.5622 & 1.565  & 1.5288 & 1.5642 &\\ \cline{2-7}
\multicolumn{1}{ |c  }{} &
\multicolumn{1}{ |c| }{Avg. correction} & 1.6004          & 1.5643 & 1.5636 & 1.5365 & 1.5722 &\\ \cline{2-7}
\multicolumn{1}{ |c  }{} &
\multicolumn{1}{ |c| }{Clm. correction} & 1.6666          & 1.5962 & 1.585  & 1.5525 & 1.5881 &\\ \cline{1-7}
\multicolumn{1}{ |c  }{\multirow{3}{*}{EWA} } &       
\multicolumn{1}{ |c| }{No correction} & 1.6264          & 1.5708 & 1.5686 & 1.531  & 1.5684 &\\ \cline{2-7}
\multicolumn{1}{ |c  }{} &
\multicolumn{1}{ |c| }{Avg. correction} & 1.6251          & 1.5712 & 1.5685 & 1.5324 & 1.57  & \\ \cline{2-7}
\multicolumn{1}{ |c  }{} &
\multicolumn{1}{ |c| }{Clm. correction} & 1.6382          & 1.5741 & 1.57   & 1.5347 & 1.5718 &\\ \cline{1-7}
\multicolumn{1}{ |c  }{\multirow{3}{*}{LAA} } &      
\multicolumn{1}{ |c| }{No correction}  & 1.6016          & 1.5622 & 1.5634 & 1.5305 & 1.569  &\\ \cline{2-7}
\multicolumn{1}{ |c  }{} &
\multicolumn{1}{ |c| }{Avg. correction} & 1.6026          & 1.5629 & 1.5636 & 1.53   & 1.5679 &\\ \cline{2-7}
\multicolumn{1}{ |c  }{} &
\multicolumn{1}{ |c| }{Clm. correction} & 1.6356          & 1.5732 & 1.5693 & 1.5332 & 1.5708 &\\ \cline{1-7}
\multicolumn{1}{ |c  }{\multirow{3}{*}{REG} } &        
\multicolumn{1}{ |c| }{No correction}   & 1.682             & 1.6037    & 1.5887    & 1.5486    & 1.5819 &\\ \cline{2-7}
\multicolumn{1}{ |c  }{} &
\multicolumn{1}{ |c| }{Avg. correction} & 1.6844          & 1.605 & 1.5892 & 1.5493 & 1.5832 &\\ \cline{2-7}
\multicolumn{1}{ |c  }{} &
\multicolumn{1}{ |c| }{Clm. correction} & 1.7564             & 1.6348    & 1.6009    & 1.5623    & 1.5911    &\\ \cline{1-7}
\multicolumn{1}{ |c  }{\multirow{3}{*}{AVG} } &        
\multicolumn{1}{ |c| }{No correction}  & 1.8141          & 1.8157 & 1.8275 & 1.8284 & 1.8813 &\\ \cline{2-7}
\multicolumn{1}{ |c  }{} &
\multicolumn{1}{ |c| }{Avg. correction} & 1.6765          & 1.6683 & 1.6758 & 1.6631 & 1.7073 &\\ \cline{2-7}
\multicolumn{1}{ |c  }{} &
\multicolumn{1}{ |c| }{Clm. correction} & 1.7176          & 1.6471 & 1.641  & 1.6092 & 1.6541 &\\ \cline{1-7}
CLM        &                 & 1.6285          & 1.5719 & 1.569  & 1.5323 & 1.5692 &\\ \cline{1-7}
\end{tabular}
\end{center}
\end{table}

\begin{table}[h]
\caption{\label{table:V_RMSE} The meridional surface wind $\RMSE_{GAW}$ for the different \textit{forecasters} and bias correction methods. The climatology is included in the ensemble.}
\begin{center}
\begin{tabular}{|c|c|c|c|c|c|c|c}
\cline{1-7}
Forecaster & Bias correction & \multicolumn{5}{ c| }{Learning period} & \\ \cline{1-7}
 &  & 5 & 10 & 15 & 20 & 25 &\\ \cline{1-7}
\multicolumn{1}{ |c  }{\multirow{3}{*}{EGA} } &
\multicolumn{1}{ |c| }{No correction}  & 1.3119          & 1.2777 & 1.282           & 1.2587 & 1.2707 &\\ \cline{2-7}
\multicolumn{1}{ |c  }{} &
\multicolumn{1}{ |c| }{Avg. correction} & 1.3108          & 1.2785 & 1.2835          & 1.263  & 1.2748 &\\ \cline{2-7}
\multicolumn{1}{ |c  }{} &
\multicolumn{1}{ |c| }{Clm. correction} & 1.364           & 1.3025 & 1.3002          & 1.2769 & 1.287 &\\ \cline{1-7}
\multicolumn{1}{ |c  }{\multirow{3}{*}{EWA} } &       
\multicolumn{1}{ |c| }{No correction} & 1.3339          & 1.2848 & 1.2874          & 1.2607 & 1.2737 &\\ \cline{2-7}
\multicolumn{1}{ |c  }{} &
\multicolumn{1}{ |c| }{Avg. correction} & 1.3329          & 1.2846 & 1.2872          & 1.2616 & 1.2731 & \\ \cline{2-7}
\multicolumn{1}{ |c  }{} &
\multicolumn{1}{ |c| }{Clm. correction} & 1.3411          & 1.2874 & 1.2885          & 1.2627 & 1.2742 &\\ \cline{1-7}
\multicolumn{1}{ |c  }{\multirow{3}{*}{LAA} } &      
\multicolumn{1}{ |c| }{No correction}  & 1.3102          & 1.2762 & 1.2816          & 1.2588 & 1.2716  &\\ \cline{2-7}
\multicolumn{1}{ |c  }{} &
\multicolumn{1}{ |c| }{Avg. correction} & 1.3124          & 1.2779 & 1.2829          & 1.2594 & 1.2717 &\\ \cline{2-7}
\multicolumn{1}{ |c  }{} &
\multicolumn{1}{ |c| }{Clm. correction} & 1.3395          & 1.286  & 1.2875          & 1.2621 & 1.2738 &\\ \cline{1-7}
\multicolumn{1}{ |c  }{\multirow{3}{*}{REG} } &        
\multicolumn{1}{ |c| }{No correction}   &  1.3744             & 1.3068    & 1.3023          & 1.274  & 1.2851 &\\ \cline{2-7}
\multicolumn{1}{ |c  }{} &
\multicolumn{1}{ |c| }{Avg. correction} & 1.3753          & 1.3075 & 1.3028          & 1.2754 & 1.2861 &\\ \cline{2-7}
\multicolumn{1}{ |c  }{} &
\multicolumn{1}{ |c| }{Clm. correction} & 1.4337             & 1.3232    & 1.3111             & 1.2837    & 1.2929     &\\ \cline{1-7}
\multicolumn{1}{ |c  }{\multirow{3}{*}{AVG} } &        
\multicolumn{1}{ |c| }{No correction}  & 1.4671          & 1.4692 & 1.4776          & 1.4779 & 1.5038 &\\ \cline{2-7}
\multicolumn{1}{ |c  }{} &
\multicolumn{1}{ |c| }{Avg. correction} & 1.3774          & 1.3732 & 1.3804          & 1.3735 & 1.3935 &\\ \cline{2-7}
\multicolumn{1}{ |c  }{} &
\multicolumn{1}{ |c| }{Clm. correction} & 1.3996          & 1.3427 & 1.34            & 1.3188 & 1.3347&\\ \cline{1-7}
CLM        &                 & 1.3352          & 1.2853 & 1.2876          & 1.2617 & 1.2731 &\\ \cline{1-7}
\end{tabular}
\end{center}
\end{table}

\begin{table}[h]
\caption{\label{table:P_RMSE} The surface pressure $\RMSE_{GAW}$ for the different \textit{forecasters} and bias correction methods. The climatology is included in the ensemble.}
\begin{center}
\begin{tabular}{|c|c|c|c|c|c|c|c}
\cline{1-7}
Forecaster & Bias correction & \multicolumn{5}{ c| }{Learning period} & \\ \cline{1-7}
 &  & 5 & 10 & 15 & 20 & 25 &\\ \cline{1-7}
\multicolumn{1}{ |c  }{\multirow{3}{*}{EGA} } &
\multicolumn{1}{ |c| }{No correction}  & 2.5682 & 2.465  & 2.4716 & 2.4189 & 2.4999 &\\ \cline{2-7}
\multicolumn{1}{ |c  }{} &
\multicolumn{1}{ |c| }{Avg. correction} &2.5228 & 2.4518 & 2.4667 & 2.412  & 2.5131 &\\ \cline{2-7}
\multicolumn{1}{ |c  }{} &
\multicolumn{1}{ |c| }{Clm. correction} & 2.6355 & 2.4967 & 2.4986 & 2.4404 & 2.5374&\\ \cline{1-7}
\multicolumn{1}{ |c  }{\multirow{3}{*}{EWA} } &       
\multicolumn{1}{ |c| }{No correction} & 2.5735 & 2.453  & 2.4607 & 2.3942 & 2.4938 &\\ \cline{2-7}
\multicolumn{1}{ |c  }{} &
\multicolumn{1}{ |c| }{Avg. correction} & 2.5712 & 2.4528 & 2.4603 & 2.3942 & 2.4942 & \\ \cline{2-7}
\multicolumn{1}{ |c  }{} &
\multicolumn{1}{ |c| }{Clm. correction} & 2.5833 & 2.4569 & 2.4631 & 2.3978 & 2.4976 &\\ \cline{1-7}
\multicolumn{1}{ |c  }{\multirow{3}{*}{LAA} } &      
\multicolumn{1}{ |c| }{No correction}  & 2.6987 & 2.5417 & 2.5194 & 2.4457 & 2.536   &\\ \cline{2-7}
\multicolumn{1}{ |c  }{} &
\multicolumn{1}{ |c| }{Avg. correction} & 2.5318 & 2.4412 & 2.454  & 2.3912 & 2.4932 &\\ \cline{2-7}
\multicolumn{1}{ |c  }{} &
\multicolumn{1}{ |c| }{Clm. correction} & 2.5821 & 2.4552 & 2.4614 & 2.3951 & 2.4957&\\ \cline{1-7}
\multicolumn{1}{ |c  }{\multirow{3}{*}{REG} } &        
\multicolumn{1}{ |c| }{No correction}   &  2.7339 & 2.5263 & 2.5076 & 2.4396 & 2.5213 &\\ \cline{2-7}
\multicolumn{1}{ |c  }{} &
\multicolumn{1}{ |c| }{Avg. correction} & 2.7338 & 2.5263 & 2.5076 & 2.4396 & 2.5213 &\\ \cline{2-7}
\multicolumn{1}{ |c  }{} &
\multicolumn{1}{ |c| }{Clm. correction} & 2.8891 & 2.5684 & 2.5303 & 2.4565 & 2.5325     &\\ \cline{1-7}
\multicolumn{1}{ |c  }{\multirow{3}{*}{AVG} } &        
\multicolumn{1}{ |c| }{No correction}  & 5.9201 & 5.917  & 5.9363 & 5.9402 & 6.0273&\\ \cline{2-7}
\multicolumn{1}{ |c  }{} &
\multicolumn{1}{ |c| }{Avg. correction} & 2.6251 & 2.6026 & 2.6277 & 2.6045 & 2.7222 &\\ \cline{2-7}
\multicolumn{1}{ |c  }{} &
\multicolumn{1}{ |c| }{Clm. correction} & 2.7146 & 2.5816 & 2.5876 & 2.5354 & 2.6441&\\ \cline{1-7}
CLM        &                 &2.5746 & 2.4531 & 2.4606 & 2.3935 & 2.4933 &\\ \cline{1-7}
\end{tabular}
\end{center}
\end{table}

\begin{table}[ht]
\centering
\caption{\label{table:T_STD_nclm} The surface temperature $\STD_{GAW}$ for the different \textit{forecasters} and bias correction methods.}
\begin{tabular}{|c|c|c|c|c|c|c|}
\hline
Forecaster           & Bias correction & \multicolumn{5}{c|}{Learning period}           \\ \hline
                     &                 & 5      & 10      & 15      & 20      & 25      \\ \hline
\multirow{3}{*}{EGA} & No correction   & 1.5722 & 1.5516  & 1.5509  & 1.5003  & 1.4699  \\ \cline{2-7} 
                     & Avg. correction & 1.2071 & 1.1681  & 1.1419  & 1.1     & 1.0898  \\ \cline{2-7} 
                     & Clm. correction & 1.1107 & 1.047   & 1.0066  & 0.97441 & 0.96212 \\ \hline
\multirow{3}{*}{EWA} & No correction   & 1.1803 & 1.1255  & 1.0967  & 1.0734  & 1.0492  \\ \cline{2-7} 
                     & Avg. correction & 1.0505 & 1.0143  & 0.99385 & 0.96995 & 0.96079 \\ \cline{2-7} 
                     & Clm. correction & 1.0001 & 0.9584  & 0.9334  & 0.9094  & 0.90286 \\ \hline
\multirow{3}{*}{LAA} & No correction   & 1.3797 & 1.1943  & 1.0777  & 0.9866  & 0.91267 \\ \cline{2-7} 
                     & Avg. correction & 1.1392 & 1.0059  & 0.91539 & 0.83884 & 0.77893 \\ \cline{2-7} 
                     & Clm. correction & 1.0596 & 0.92525 & 0.83163 & 0.77086 & 0.71868 \\ \hline
\multirow{3}{*}{AVR} & No correction   & 1.7773 & 1.7817  & 1.7849  & 1.7888  & 1.7888  \\ \cline{2-7} 
                     & Year bias       & 1.2573 & 1.2335  & 1.2129  & 1.2     & 1.1874  \\ \cline{2-7} 
                     & Month bias      & 1.1132 & 1.053   & 1.0151  & 0.99245 & 0.9729  \\ \hline
\end{tabular}
\end{table}

\begin{table}[ht]
\centering
\caption{\label{table:U_STD_nclm} The zonal surface wind $\STD_{GAW}$ for the different \textit{forecasters} and bias correction methods.}
\begin{tabular}{|c|c|c|c|c|c|c|}
\hline
Forecaster           & Bias correction & \multicolumn{5}{c|}{Learning period}           \\ \hline
                     &                 & 5      & 10      & 15      & 20      & 25      \\ \hline
\multirow{3}{*}{EGA} & No correction   & 1.3211 & 1.3083  & 1.2938  & 1.2958  & 1.301   \\ \cline{2-7} 
                     & Avg. correction & 1.218  & 1.1991  & 1.1895  & 1.1955  & 1.1906  \\ \cline{2-7} 
                     & Clm. correction & 1.1857 & 1.1267  & 1.1117  & 1.1121  & 1.1013  \\ \hline
\multirow{3}{*}{EWA} & No correction   & 1.1648 & 1.1222  & 1.1094  & 1.1066  & 1.1125  \\ \cline{2-7} 
                     & Avg. correction & 1.1275 & 1.0924  & 1.0866  & 1.0938  & 1.0903  \\ \cline{2-7} 
                     & Clm. correction & 1.0902 & 1.034   & 1.0298  & 1.0247  & 1.0193  \\ \hline
\multirow{3}{*}{LAA} & No correction   & 1.2084 & 1.0821  & 0.98004 & 0.91758 & 0.86497 \\ \cline{2-7} 
                     & Avg. correction & 1.1193 & 1.0126  & 0.92726 & 0.88286 & 0.83293 \\ \cline{2-7} 
                     & Clm. correction & 1.0989 & 0.96991 & 0.88354 & 0.83389 & 0.7855  \\ \hline
\multirow{3}{*}{AVR} & No correction   & 1.3665 & 1.3673  & 1.3694  & 1.3751  & 1.377   \\ \cline{2-7} 
                     & Avg. correction & 1.1962 & 1.189   & 1.1858  & 1.1892  & 1.187   \\ \cline{2-7} 
                     & Clm. correction & 1.1566 & 1.1034  & 1.0871  & 1.085   & 1.0759  \\ \hline
\end{tabular}
\end{table}

\begin{table}[ht]
\centering
\caption{\label{table:V_STD_nclm} The meridional surface wind $\STD_{GAW}$ for the different \textit{forecasters} and bias correction methods.}
\begin{tabular}{|c|c|c|c|c|c|c|}
\hline
Forecaster           & Bias correction & \multicolumn{5}{c|}{Learning period}           \\ \hline
                     &                 & 5               & 10      & 15      & 20      & 25      \\ \hline
\multirow{3}{*}{EGA} & No correction   & 1.0428          & 1.0292  & 1.0279  & 1.0221  & 1.0122  \\ \cline{2-7} 
                     & Avg. correction & 0.959           & 0.95016 & 0.94543 & 0.94398 & 0.93683 \\ \cline{2-7} 
                     & Clm. correction & 0.93743         & 0.9019  & 0.88682 & 0.88115 & 0.86831 \\ \hline
\multirow{3}{*}{EWA} & No correction   & 0.91457         & 0.88655 & 0.87795 & 0.87873 & 0.86699 \\ \cline{2-7} 
                     & Avg. correction & 0.88297         & 0.87172 & 0.86361 & 0.86215 & 0.85405 \\ \cline{2-7} 
                     & Clm. correction & 0.85492         & 0.83007 & 0.81715 & 0.80731 & 0.79725 \\ \hline
\multirow{3}{*}{LAA} & No correction   & 0.94936         & 0.85194 & 0.78035 & 0.7294  & 0.67994 \\ \cline{2-7} 
                     & Avg. correction & 0.88457         & 0.80834 & 0.74702 & 0.70583 & 0.66468 \\ \cline{2-7} 
                     & Clm. correction & 0.86813         & 0.76843 & 0.69957 & 0.65128 & 0.60696 \\ \hline
\multirow{3}{*}{AVR} & No correction   & 1.0623          & 1.0622  & 1.0646  & 1.0653  & 1.0663  \\ \cline{2-7} 
                     & Avg. correction & 0.9454          & 0.93813 & 0.93673 & 0.93579 & 0.9352  \\ \cline{2-7} 
                     & Clm. correction & 0.92446         & 0.88248 & 0.86953 & 0.86301 & 0.85696 \\ \hline
\end{tabular}
\end{table}

\begin{table}[ht]
\centering
\caption{\label{table:P_STD_nclm} The surface pressure $\STD_{GAW}$ for the different \textit{forecasters} and bias correction methods.}
\begin{tabular}{|c|c|c|c|c|c|c|}
\hline
Forecaster           & Bias correction & \multicolumn{5}{c|}{Learning period}       \\ \hline
                     &                 & 5      & 10     & 15     & 20     & 25     \\ \hline
\multirow{3}{*}{EGA} & No correction   & 4.4105 & 4.351  & 4.3096 & 4.2797 & 4.2831 \\ \cline{2-7} 
                     & Avg. correction & 2.4633 & 2.4556 & 2.451  & 2.453  & 2.4329 \\ \cline{2-7} 
                     & Clm. correction & 2.4548 & 2.3638 & 2.3503 & 2.343  & 2.314  \\ \hline
\multirow{3}{*}{EWA} & No correction   & 2.5294 & 2.3962 & 2.3722 & 2.3379 & 2.327  \\ \cline{2-7} 
                     & Avg. correction & 2.3429 & 2.3491 & 2.3509 & 2.3472 & 2.3384 \\ \cline{2-7} 
                     & Clm. correction & 2.3384 & 2.3005 & 2.3073 & 2.2899 & 2.2663 \\ \hline
\multirow{3}{*}{LAA} & No correction   & 3.7348 & 3.267  & 2.9498 & 2.7572 & 2.5945 \\ \cline{2-7} 
                     & Avg. correction & 2.2853 & 2.1309 & 2.0012 & 1.9073 & 1.8249 \\ \cline{2-7} 
                     & Clm. correction & 2.2975 & 2.0974 & 1.9856 & 1.8839 & 1.7998 \\ \hline
\multirow{3}{*}{AVR} & No correction   & 5.8832 & 5.885  & 5.8841 & 5.8911 & 5.8787 \\ \cline{2-7} 
                     & Avg. correction & 2.4017 & 2.3929 & 2.3839 & 2.3834 & 2.3636 \\ \cline{2-7} 
                     & Clm. correction & 2.3771 & 2.2727 & 2.2459 & 2.2339 & 2.2015 \\ \hline
\end{tabular}
\end{table}

\begin{table}[ht]
\centering
\caption{\label{table:T_STD} The surface temperature $\STD_{GAW}$ for the different \textit{\textit{forecasters}} and bias correction methods. The climatology is included in the ensemble.}
\begin{tabular}{|c|c|c|c|c|c|c|}
\hline
Forecaster           & Bias correction & \multicolumn{5}{c|}{Learning period}                 \\ \hline
                     &                 & 5        & 10       & 15       & 20       & 25       \\ \hline
\multirow{3}{*}{EGA} & No correction   & 0.85636  & 0.79338  & 0.74092  & 0.74888  & 0.77041  \\ \cline{2-7} 
                     & Avg. correction & 0.69405  & 0.65224  & 0.60268  & 0.60546  & 0.59947  \\ \cline{2-7} 
                     & Clm. correction & 0.82139  & 0.76932  & 0.72812  & 0.69949  & 0.68463  \\ \hline
\multirow{3}{*}{EWA} & No correction   & 0.069617 & 0.04385  & 0.034291 & 0.034327 & 0.046811 \\ \cline{2-7} 
                     & Avg. correction & 0.093904 & 0.055604 & 0.042137 & 0.042974 & 0.0541   \\ \cline{2-7} 
                     & Clm. correction & 0.32004  & 0.22664  & 0.18245  & 0.18477  & 0.22204  \\ \hline
\multirow{3}{*}{LAA} & No correction   & 0.86891  & 0.69606  & 0.60843  & 0.56257  & 0.52853  \\ \cline{2-7} 
                     & Avg. correction & 0.64118  & 0.49176  & 0.41823  & 0.37824  & 0.35085  \\ \cline{2-7} 
                     & Clm. correction & 0.61502  & 0.44528  & 0.35476  & 0.31043  & 0.28258  \\ \hline
\multirow{3}{*}{AVR} & No correction   & 1.8474   & 1.8482   & 1.8513   & 1.8555   & 1.8556   \\ \cline{2-7} 
                     & Avg. correction & 1.301    & 1.2736   & 1.2536   & 1.2398   & 1.2267   \\ \cline{2-7} 
                     & Clm. correction & 1.1244   & 1.0643   & 1.0281   & 1.006    & 0.98709  \\ \hline
\end{tabular}
\end{table}

\begin{table}[ht]
\centering
\caption{\label{table:U_STD} The zonal surface wind $\STD_{GAW}$ for the different \textit{forecasters} and bias correction methods. The climatology is included in the ensemble.}
\label{my-label}
\begin{tabular}{|c|c|c|c|c|c|c|}
\hline
Forecaster           & Bias correction & \multicolumn{5}{c|}{Learning period}                 \\ \hline
                     &                 & 5        & 10       & 15       & 20       & 25       \\ \hline
\multirow{3}{*}{EGA} & No correction   & 0.73276  & 0.65502  & 0.60472  & 0.6103   & 0.55895  \\ \cline{2-7} 
                     & Avg. correction & 0.73511  & 0.66343  & 0.60299  & 0.60858  & 0.55592  \\ \cline{2-7} 
                     & Clm. correction & 0.84851  & 0.76309  & 0.71594  & 0.69938  & 0.65361  \\ \hline
\multirow{3}{*}{EWA} & No correction   & 0.084994 & 0.046667 & 0.035807 & 0.034519 & 0.030238 \\ \cline{2-7} 
                     & Avg. correction & 0.11876  & 0.066557 & 0.052488 & 0.046515 & 0.041827 \\ \cline{2-7} 
                     & Clm. correction & 0.34414  & 0.19312  & 0.13497  & 0.11729  & 0.10056  \\ \hline
\multirow{3}{*}{LAA} & No correction   & 0.70173  & 0.54873  & 0.47246  & 0.43713  & 0.41112  \\ \cline{2-7} 
                     & Avg. correction & 0.60517  & 0.46204  & 0.39078  & 0.35748  & 0.33391  \\ \cline{2-7} 
                     & Clm. correction & 0.60982  & 0.43764  & 0.35328  & 0.3177   & 0.29133  \\ \hline
\multirow{3}{*}{AVR} & No correction   & 1.439    & 1.4335   & 1.4333   & 1.4364   & 1.4383   \\ \cline{2-7} 
                     & Avg. correction & 1.2418   & 1.2278   & 1.2233   & 1.2237   & 1.2214   \\ \cline{2-7} 
                     & Clm. correction & 1.173    & 1.1189   & 1.1023   & 1.0993   & 1.0906   \\ \hline
\end{tabular}
\end{table}

\begin{table}[ht]
\centering
\caption{\label{table:V_STD} The meridional surface wind $\STD_{GAW}$ for the different \textit{forecasters} and bias correction methods. The climatology is included in the ensemble.}
\begin{tabular}{|c|c|c|c|c|c|c|}
\hline
Forecaster           & Bias correction & \multicolumn{5}{c|}{Learning period}                 \\ \hline
                     &                 & 5        & 10       & 15       & 20       & 25       \\ \hline
\multirow{3}{*}{EGA} & No correction   & 0.56695  & 0.50983  & 0.47015  & 0.47185  & 0.44052  \\ \cline{2-7} 
                     & Avg. correction & 0.56347  & 0.52115  & 0.47243  & 0.47293  & 0.43637  \\ \cline{2-7} 
                     & Clm. correction & 0.67315  & 0.61536  & 0.56948  & 0.55546  & 0.52289  \\ \hline
\multirow{3}{*}{EWA} & No correction   & 0.056521 & 0.031311 & 0.025234 & 0.025545 & 0.022147 \\ \cline{2-7} 
                     & Avg. correction & 0.078621 & 0.046682 & 0.036763 & 0.030673 & 0.026266 \\ \cline{2-7} 
                     & Clm. correction & 0.26654  & 0.15412  & 0.11447  & 0.094416 & 0.080736 \\ \hline
\multirow{3}{*}{LAA} & No correction   & 0.54687  & 0.42964  & 0.37097  & 0.3416   & 0.32204  \\ \cline{2-7} 
                     & Avg. correction & 0.48228  & 0.36769  & 0.31136  & 0.28407  & 0.26622  \\ \cline{2-7} 
                     & Clm. correction & 0.47982  & 0.34436  & 0.27938  & 0.25008  & 0.22985  \\ \hline
\multirow{3}{*}{AVR} & No correction   & 1.1258   & 1.119    & 1.1194   & 1.1178   & 1.1187   \\ \cline{2-7} 
                     & Avg. correction & 0.98815  & 0.9738   & 0.97075  & 0.96766  & 0.96716  \\ \cline{2-7} 
                     & Clm. correction & 0.93731  & 0.89451  & 0.88127  & 0.87434  & 0.86855  \\ \hline
\end{tabular}
\end{table}

\begin{table}[ht]
\centering
\caption{\label{table:P_STD} The surface pressure $\STD_{GAW}$ for the different \textit{forecasters} and bias correction methods. The climatology is included in the ensemble.}
\begin{tabular}{|c|c|c|c|c|c|c|}
\hline
Forecaster           & Bias correction & \multicolumn{5}{c|}{Learning period}                \\ \hline
                     &                 & 5       & 10       & 15       & 20       & 25       \\ \hline
\multirow{3}{*}{EGA} & No correction   & 2.9392  & 2.717    & 2.5768   & 2.5104   & 2.3833   \\ \cline{2-7} 
                     & Avg. correction & 1.5787  & 1.5048   & 1.4308   & 1.3989   & 1.2599   \\ \cline{2-7} 
                     & Clm. correction & 1.7647  & 1.631    & 1.5915   & 1.5789   & 1.4751   \\ \hline
\multirow{3}{*}{EWA} & No correction   & 0.18324 & 0.099206 & 0.073743 & 0.064644 & 0.055238 \\ \cline{2-7} 
                     & Avg. correction & 0.19148 & 0.13322  & 0.10974  & 0.088419 & 0.07355  \\ \cline{2-7} 
                     & Clm. correction & 0.60822 & 0.39215  & 0.31246  & 0.27342  & 0.22431  \\ \hline
\multirow{3}{*}{LAA} & No correction   & 2.9241  & 2.4551   & 2.1702   & 2.0292   & 1.9133   \\ \cline{2-7} 
                     & Avg. correction & 1.1266  & 0.85932  & 0.71971  & 0.65848  & 0.60953  \\ \cline{2-7} 
                     & Clm. correction & 1.1816  & 0.83857  & 0.67345  & 0.60363  & 0.54823  \\ \hline
\multirow{3}{*}{AVR} & No correction   & 6.2919  & 6.2863   & 6.283    & 6.2882   & 6.2762   \\ \cline{2-7} 
                     & Avg. correction & 2.4483  & 2.4306   & 2.42     & 2.4176   & 2.3972   \\ \cline{2-7} 
                     & Clm. correction & 2.3963  & 2.2911   & 2.2639   & 2.2515   & 2.2187   \\ \hline
\end{tabular}
\end{table}

\end{document}